\documentstyle[prl,aps]{revtex}

\begin{document}

\draft
\title{Hole Solutions in the 1d Complex Ginzburg-Landau Equation}
\author{Stefan Popp$^1$, Olaf Stiller$^1$, Igor~Aranson$^2$
        and Lorenz Kramer$^1$}
\address{$^1$ Physikalisches Institut der Universit\"at Bayreuth,
         D-95440 Bayreuth, Germany \\
         $^2$ Department of Physics, Bar-Ilan University,
         Ramat Gan 52900, Israel}
\date{\today}
\maketitle

\begin{abstract}
The cubic Complex Ginzburg-Landau Equation (CGLE)
has a one parameter family of traveling localized
source solutions.
These so called 'Nozaki-Bekki holes'
are (dynamically) stable in some parameter range,
but always structually unstable:
A perturbation of the
equation in general leads to a
(positive or negative) monotonic acceleration or
an oscillation of the holes.
This confirms that the cubic CGLE has an inner symmetry.
As a consequence small perturbations change some of the qualitative
dynamics of the cubic CGLE and enhance or suppress
spatio-temporal intermittency in some parameter range.
An analytic stability analysis of holes in the cubic
CGLE and a semianalytical treatment of the acceleration
instability in the perturbed equation is performed by
using matching and perturbation methods.
Furthermore we treat the asymptotic hole--shock interaction.
The results, which can be obtained fully analytically in the
nonlinear Schroedinger limit,
are also used for the quantitative description
of modulated solutions made up of periodic arrangements
of traveling holes and shocks.
\end{abstract}

\pacs{PACS numbers: 47.20.-k; 05.45.+b}

(Submitted to Physica D)

\section{Introduction}
Spatially extended oscillatory media in the weakly nonlinear regime
can be described by a complex Ginzburg Landau equation (CGLE)
which may be derived from an amplitude and slow-variation expansion
\cite{CGL}. In 1d it reads
\begin{eqnarray}
  \label{eq1}
  \partial_t A = \left[1 + (1+ib) \partial_x^2 - (1+ic)|A|^2
  + d|A|^4 \right] A
\end{eqnarray}
where  $b$ and $c$ are real constants .
Length, time and the complex order parameter $A$ have been scaled
in the usual way for the cubic CGLE.
In eq.(\ref{eq1}) we have included
a quintic term with a small complex prefactor
$ d = d' + i d ''$ ($ |d| \ll 1 $).
We will refer to eq.(\ref{eq1})
with $d \neq 0$ as the "perturbed (cubic) CGLE"
and treat the $d $ term only perturbatively to lowest order
$O(d)$.

As already pointed out in reference \cite{PRL}
arbitrarily small $|d| $ may change the situation drastically,
and eq.(\ref{eq1}) without fifth order term ($d \equiv 0$)
is not sufficient to describe the qualitative dynamics of the
system in the parameter regime considered here, i.e. where Nozaki-Bekki  hole
solutions (see below)
influence the dynamics.
{}From the derivation of the CGLE one expects
in a real physical system
$d \propto \epsilon$
where $\epsilon $ is the distance from threshold.
There are other corrections to the cubic CGLE
which should have the
same order of magnitude but
their "perturbative effect" is expected to be rather similar and we have
chosen the $d$ term in eq.(\ref{eq1}) to represent "higher-order
perturbations".
While in reference \cite{PRL} we presented a phenomenological
theory for phenomena observed numerically,
in this work we will determine
the important quantities of the theory with the help of
analytical or semi-analytical methods.

Equation (\ref{eq1}) has the family of traveling-wave solutions
\begin{eqnarray}
  \label{plane}
  A = R_q
  e^{i(q x- \omega(q) t)}
  \quad\quad
  \mbox{with}
  \quad
  R_q= \sqrt{1-q^2 + d' (1-q^2)^2}
  \; ,
  \quad
  0 \leq q^2 \leq 1
\end{eqnarray}
and the dispersion relation
\begin{eqnarray}
  \label{eq2}
  \omega(q) = c R_q^2 +bq^2 -  d'' R_q^4
  \;.
\end{eqnarray}
(As mentioned above all formulas are valid only at order  $d^1$.)

\subsection{Localized Solutions}

Other important solutions are localized objects $A_v$
moving  with constant velocity $v$ connecting
asymptotic plane-wave states.
In the co--moving frame ($\zeta = x-vt$) the
plane-wave dispersion relation
reads $\Omega(q,v)=\omega(q)-vq$.
Dealing with such constantly moving solutions it is convenient
to write the CGLE (\ref{eq1}) in a
co--moving frame rotating with constant frequency $\Omega$
\begin{eqnarray}
  \partial_t A(\zeta,t) &=& F_{v,\Omega}[A]
  \nonumber\\
  & := &
  \left(1+i\Omega + (1+ib) \partial_{\zeta}^2 + v  \partial_{\zeta}
  - (1+ic)|A|^2  + d |A|^4 \right) A
  \label{eq1frame}
\end{eqnarray}
Clearly, in order to ensure phase conservation,
it is necessary that
in this co--moving frame the plane waves on both sides rotate
with the same frequency,
i.e. $\omega(q_i)-vq_i= \Omega$ for $i=1,2$.
{}From this and eqs.(\ref{plane}) and (\ref{eq2})
one can derive the relation:
\begin{eqnarray}
  \label{topol}
  v =\frac{\omega(q_2)-\omega(q_1)}{q_2-q_1}
  = 2(b-c) k - 2 (c d'- d'')
  (R_{q_1}^2 + R_{q_2}^2)
   k
  \\
  \mbox{with}\qquad  k=\frac{1}{2} (q_1+q_2)
\end{eqnarray}
Note that in the cubic CGLE this yields $v= 2(b-c) k$
which is the mean of the group velocity $\partial_q \omega(q)$
of the asymptotic plane waves.
One differentiates between sources and sinks (often called shocks)
depending on whether the group velocity (and thus causality)
points inward or outward in the co-moving frame.
Thus sources should determine
the behavior of the asymptotic states
while shocks are passive and play only a minor role.
So in the following we will focus mainly on the investigation of
sources.

For a standing ($v=0$) (antisymmetric) source solution
eq.(\ref{topol})
is fulfilled trivially due to symmetry ($q_1=-q_2$).
For $v\neq 0$, however, eq.(\ref{topol})
is not trivial at all and,
as pointed out by
van Saarloos and Hohenberg \cite{HB},
the problem to find such
source solutions of eq.(\ref{eq1}) is in
general overdetermined. Using counting arguments they showed that
from a systematic point of view
only standing sources should exist.
Nevertheless the cubic CGLE (eq.(1) with $ d = 0$)
has an analytic one-parameter family of sources with different
velocities $v$, the "hole solutions" of Nozaki and Bekki \cite{NB}
(NB holes).
They can be written in the form
\begin{eqnarray}
  \label{NB}
  A^{NB}_v & := & Z_v \exp(i\chi_v - i\Omega t)
   \nonumber \\
    &  = &
           [\hat B \partial_{\zeta}\varphi_v(\kappa\zeta) +\hat A v] \times
           \exp[i \varphi_v(\kappa\zeta) + i\hat  k v \zeta - i\Omega t]
  \\ \mbox{where}&&\quad \varphi_v(\kappa\zeta)
  = \hat\kappa^{-1}\ln \cosh(\kappa\zeta)
  \; .
  \nonumber
\end{eqnarray}
Symbols with a "hat" denote constants depending only on  $b$
and $c$, e.g. $\hat  k = 1/(2(\!b\!-\!c\!))$ .
The frequency $\Omega$ and $\kappa^2$ are linear functions of
$v^2$.
With exception of $\hat A $
all coefficients are real.
The asymptotic plane waves for $\zeta\to \pm\infty$
have wavenumbers
\begin{eqnarray}
\label{q}
  q_{1/2}=  k \pm K
  \;
  \qquad \mbox{with} \;\;
   K = \kappa/\hat\kappa
  \; , \quad
   k = v\hat k
  \; .
\end{eqnarray}
$\kappa^2$ becomes zero at a maximal velocity $\pm v_{max}$, and
here the hole solution merges
with a plane wave with wavenumber
$q_1=q_2 = v_{max}/(2(\!b\!-\!c\!))$.
The relations are completely derived in Appendix \ref{NB_a}.
The resulting algebraic equations (8 equations for 8 parameters)
yield the one parameter family.
The derivation in Appendix \ref{NB_a}
demonstrates the necessary dependence in a
transparent way.

The NB holes are connected with the 3-parameter family of
dark soliton solutions that exist in the limit
$b=\pm\infty,\ c=\pm\infty$ , where eq.(\ref{eq1})
reduces to the defocussing nonlinear Schroedinger equation
(see next paper),
and with the one-parameter family of static
(all velocities collapse to zero)
saddle-point solutions of the nonlinear diffusion equation
obtained in the limit $b=c=0$ \cite{LA}.
Then one has a potential and the continuous family can
be associated with (global) gauge invariance through
Noether's theorem.
Clearly then perturbations of the equation that preserve
gauge invariance and the potential property,
like a quintic term with a real prefactor,
do not destroy the family.

An important result of this article is that in general
the family of hole solutions is
destroyed by the higher-order perturbation ($ d\neq 0$)
leaving only the standing hole.
Thus moving NB holes are structurally unstable.

\subsection{Stability and Destruction of Hole Solutions}
\label{review}
The stability of the hole solutions (for $ d = 0$)
has first been investigated by Sakaguchi \cite{S85}
in direct simulations of the CGLE.
The perturbational equation of the CGLE (\ref{eq1frame})
with $A=A_v + {\cal W}$ describing the stability problem
\begin{eqnarray}
  \label{stab}
  \partial_t {\cal W} = {\cal L}_{v} {\cal W}
  \qquad \mbox{with}\quad {\cal L}_v{\cal W}
  := \frac{\delta F_{v,\Omega}} {\delta A}[A_v] \,
          {\cal W}  \;
  + \;
  \frac{\delta F_{v,\Omega}}{\delta A^*} [A_v]
         {\cal W}^*  \;
\end{eqnarray}
(where now $A_v=A^{NB}_v$ and $\Omega=\Omega(v)$)
describing the stability problem was then studied
by Chat\'e and Manneville numerically for $v=0,v=0.2$
\cite{MC}
and by Sasa and Iwamoto
semianalytically for $v=0$ \cite{sasa}.
As a result hole solutions were found to be stable in
a narrow region of the $b$-$c$ plane which is
shown in Fig.1 for the standing hole ($v=0$) .
{}From below the region is bounded
by the border of (absolute) stability of the
emitted plane waves with wavenumber $q(b,c)$ (see eq.(\ref{NB}))
corresponding to the continuous spectrum of $ {\cal L}_{v} $.
{}From the other sides the stable range is bounded by
the instability of the core with respect to localized
eigenmodes corresponding to a discrete spectrum of $ {\cal L}_{v} $.
The core instability turns out to be
connected with a stationary bifurcation where the destabilizing mode
at threshold passes through
the neutral mode $ {\cal W} = {\mit \Phi}_{tr}$
(see eq.(\ref{neutral}))
of ${\cal L}_{v}$ which can be derived from
translational invariance of the CGLE
\cite{MC,sasa}.
The cores of moving holes were found to be more stable
than those of the standing ones \cite{MC}.

Now we turn to the situation with small but finite $  d $.
Our simulations show that in the stable range
moving holes are then in general either accelerated and eventually
destroyed or slowed down and stopped to the standing
hole solution depending on the phase of $ d$.
In particular for real $  d =  d' $ one has
\begin{eqnarray}
\label{slowed}
\left[\frac{\partial_t v}{v}\right]
  {>\atop <}  0 \Leftrightarrow   d' {>\atop <} 0
\end{eqnarray}
One finds that the Nozaki-Bekki hole relations connecting
the core velocity and the emitted wavenumbers
(see eq.(\ref{NB}) and App.\ref{NB_a})
are (almost) satisfied at each instant
during the acceleration process.
The acceleration thus occurs approximately
along the NB hole family
and it can be described by taking  $ v = v(t) $ as a slowly
varying variable while
other degrees of freedom follow adiabatically.

{}From these numerical observations we conclude that for
arbitrarily fixed values of $ b$,$c$ and varying $ d$
the standing hole solution
undergoes a symmetry-breaking stationary bifurcation at $  d = 0 $.
The (real) growth rate $ \lambda_{ d} $ of the unstable mode is
proportional to the acceleration
\begin{eqnarray}
 \label{lambd}
 \lambda_{ d} =\lim_{v\rightarrow 0} \dot v/v \,\, .
\end{eqnarray}
We will refer to this instability as the "acceleration instability"
while the term "core instability" will be used for the bifurcation
where $b$ and/or $c$ is changed while $ d$ is kept constant
(including the case $  d = 0 $).

Of special interest is the case where the core-stability line
is crossed with  $  d \neq 0 $
(which actually corresponds to the
physically relevant situation).
Then the two modes which cause the acceleration instability
and the (stationary) core instability (with $  d \equiv 0 $)
are coupled which leads to a Hopf bifurcation.
(nur bei $d$ stabilisierend)
As we showed in \cite{PRL} the normal form for this bifurcation
-- valid for small $ d,v,u $ -- is
     \begin{eqnarray}   \label{phen}
\begin{array}{ccccc}
\dot{u} &=& (\lambda - s v^2)u &+& d_1 v     \\
\dot{v} &=&  \mu u             &+& d_2 v \quad .
\end{array}
     \end{eqnarray}
Here $u$ and $v$ are the amplitudes of the
core-instability and acceleration-instability modes respectively.
$s$ and $\mu$ are of order $d^0$
while $d_{1,2}$ must be of order $d^1$ since in the absence
of a perturbation holes with nonzero velocity exist.
For $d=0$ the parameter $\lambda$ can be identified
with the  growth rate  of the core-instability mode.
(In the general case $\lambda$ has a correction  proportional to
$d$ causing a shift of the threshold.)
The nonlinear term in eq.(\ref{phen}) takes care of the
fact that moving holes are more stable
than standing ones and at the same time saturates the instability
($s>0$).

Far away from the core-instability threshold where $\lambda$ is
strongly negative
$u$ can be eliminated adiabatically from eqs.(\ref{phen})
which yields
\begin{eqnarray}
  \label{phen_a}
  \dot{v} = (d_2 - d_1\frac{\mu}{\lambda}){v}
\end{eqnarray}
The term in brackets can be identified as $\lambda_{d}$
from eq.(\ref{lambd}).

After introducing our general tools and concepts
in Sec.\ref{tools} we will present a detailed
perturbation analysis in Sec.\ref{pert}.
Since the core instability
(for the case  $ d = 0 $)
as well as the acceleration instability
are of the stationary type (real growth rate) we will first
investigate the stationary bifurcation scenario more closely
in Sec.\ref{stat_bif}.
Our approach allows to treat both instabilities within the framework
of the same formalism.
Then we will derive a fully analytic expression for the
core instability line in the unperturbed equation ($ d = 0 $)
(Sec.\ref{core_inst}),
present a semi-analytic method for treating the acceleration
instability in the case $ d \neq 0 $
(Sec.\ref{acc_inst})
and investigate the interaction between holes and shocks
in Sec.\ref{int_hs}.
A comparison of our analysis with simulations of the CGLE
is presented in Sec.\ref{comparison}.
Finally we will use our results for a quantitative
description of moving periodically modulated solutions
of the perturbed CGLE (Sec.\ref{per_sol}).

\section{Tools and Concepts}
\label{tools}

\subsection{The Linear Operator}
\label{lin_op}
The linear operator ${\cal L}_v$
introduced in eq.(\ref{stab})
has two bounded neutral modes
\begin{eqnarray}
\label{neutral}
{\mit\Phi}_{rot}= i A_v \qquad
{\mit\Phi}_{tr} =  \partial_{\zeta} A_v
\end{eqnarray}
related to gauge and translational invariance of the CGLE.

For many applications it is convenient to introduce a transformation
$W={\cal W}  \exp (-i \vartheta_{v})$ where $\vartheta_{v}$
is some real field (often the phase of $A_v$).
Equation (\ref{stab}) then goes over into
\begin{eqnarray}
  \label{L_eck}
  \partial_t W = L_{v} W
  :=
  \exp (- i \vartheta_{v}) {\cal L}_{v}
  \left(\exp (i \vartheta_{v}) W \right)
  \quad .
\end{eqnarray}
For the NB solutions ($d=0$)
we choose $\vartheta_{v}=\chi_v$ obtaining
\begin{eqnarray}
\label{Lv}
L^{NB}_v W & = &
   \left( (1+i\Omega) +
          (1+ib)(\partial_{\zeta} + i\partial_{\zeta}\chi_{v})^2 +
          v (\partial_{\zeta} + i\partial_{\zeta}\chi_{v}) -
          2 (1+ic) |A^{NB}_{v}|^2 \right) W -
   (1+ic) |A^{NB}_{v}|^2 W^*
   \nonumber \\
      & = &
   (1\!+\!ib) \partial_{\zeta\zeta} W \; + \;
   \left( 2 K i (1\!+\!ib)
   \tanh(\kappa\zeta) + 2 k i(1\!+\!ic) \right)
          \partial_{\zeta} W
   \nonumber \\
      & + &
   \left( (1\!+\!ic)(- K^2\hat B^2\tanh^2(\kappa\zeta) +
              2 k K\tanh(\kappa\zeta) - 2v^2\hat A^2)
          + (1\!+\!i\Omega+(1\!+\!ib)i K\kappa)/\cosh^2(\kappa\zeta)
          + iv k -  k^2(1\!+\!ib) \right) W
   \nonumber \\
      & - &
   \left( (1\!+\!ic)(v\hat A+ K \hat B\tanh(\kappa\zeta))^2 \right) W^*
   \quad .
\end{eqnarray}
Stationary instabilities are described by the eigenvalue problem
related to eq.(\ref{L_eck})
\begin{eqnarray}
  \label{eigenvalue}
  L_v  W = \lambda  W
\end{eqnarray}
with real $\lambda$.
The bounded neutral modes from eq.(\ref{neutral}) are now
\begin{eqnarray}
  \label{phi_rot}
  \Phi_{rot}   & = &  i Z_v
  \; = \;
  i (v\hat A +  K \hat B \tanh (\kappa \zeta))
  \\
  \label{phi_tr}
  \Phi_{tr} & = &
  \partial_{\zeta} Z_v + i Z_v \partial_{\zeta} \chi_v
  \; = \;
   K\hat B \kappa / \cosh^2 (\kappa \zeta) \, + \,
  \left(  k +  K \tanh (\kappa \zeta) \right) \, \Phi_{rot}
  \; .
\end{eqnarray}
The other two fundamental modes satisfying $L^{NB}_v \Phi = 0$
(irrespective of boundary conditions)
can be expressed in terms of
generalized hypergeometric functions.

Asymptotically ($  |\kappa\zeta | \gg 1 $) one has plane waves,
the operator $ L_v $
becomes space  independent and the four
fundamental modes
behave exponentially $ \sim e^{p_i\zeta} $.
The  exponents $p_i \; (i=1...4) $
can be obtained from the characteristic polynomial.
Two of the exponents can be extracted from the bounded neutral
modes $\Phi_{rot}$ and $\Phi_{tr}$
They are
\begin{eqnarray}
  \label{exponent12}
   p_1 = 0    \quad  \mbox{and} \quad   p_2 = \mp 2\kappa  \qquad
  \mbox{for}\; \zeta \to \pm \infty
\end{eqnarray}
The other - in general complex - exponents
$ p_3 $ and $ p_4 $
are given in Appendix \ref{app_pws}.
Their real part is always positive/negative for
$\zeta \stackrel{>}{<} 0$  showing that
the other two fundamental modes
are exponentially growing in space.

\subsection{Condition for the Existence of a Family}
\label{family_cond}
A necessary condition for a standing hole solution
$ A_0:=A_{v=0} $ of the CGLE
to be embedded in a continuous one-parameter family of moving hole
solutions $A_{v}$ can be obtained
by taking the derivative of the stationary CGLE
$ F_{v,\Omega}[A_{v}]=0 $ with respect to $v$ at $v=0$ :
\begin{eqnarray}
\label{fam_bed}
& {\cal L}_0 {\mit \Phi}_{fam} \left[ =
  \left(\frac{\partial F_{v,\Omega}}{\partial v} \right)_{v=0} [A_{0}]
  = \partial_{\zeta} A_{0} \right] \;
  = {\mit \Phi}_{tr} &  \\
& \mbox{with} \quad
  {\mit \Phi}_{fam} = \left.- \frac{dA_{v}}{dv} \right|_{v=0} &
  \nonumber
\end{eqnarray}
(Here the fact has been used that for
symmetry reasons $\Omega $ can only depend on $v^2$.)
Equation (\ref{fam_bed}) shows that the existence of a hole family
implies that the translation mode
$ {\mit\Phi}_{tr}=\partial_{\zeta} A_{0} $
has an inverse image ${\mit \Phi}_{fam}$
under ${\cal L}_0:= {\cal L}_{v=0} $
which for large $|\zeta|$ grows linearly in space
(${\mit \Phi}_{fam}\sim i\zeta A_v$)
corresponding to an asymptotic wavenumber change
(c.f. eq.(\ref{phi_fam})).
In the cubic case ($d = 0$) eq.(\ref{fam_bed})
with this boundary condition
is solved by inserting the Nozaki--Bekki solution
$ A_v=A^{NB}_v $.
Then instead of eq.(\ref{fam_bed}) one may write
(c.f. eqs.(\ref{L_eck},\ref{NB},\ref{q}))
\begin{eqnarray}
  \label{phi_fam}
  L^{NB}_0 \Phi_{fam} &=& \Phi_{tr} \; ,
  \quad  \mbox{with}\;\;
    - \Phi_{fam}
   \; = \;
   \hat A + i \hat  k \hat B  K \zeta \tanh (\kappa \zeta)
   \,\, .
\end{eqnarray}
In general ($ d \neq 0 $)
all solutions of eq.(\ref{fam_bed}) are found to diverge exponentially
for $\zeta \to \pm\infty$ (see below) which is consistent
with the fact that the hole family is destroyed
by the higher-order perturbation.
In the limit $ 0 \neq \left| \frac{b-c}{1+bc} \right| \ll 1 $ ,
which includes the relaxative case $b=c=0$ as well as the
conservative and fully integrable
nonlinear Schroedinger limit ($b,c\rightarrow \infty$),
we show this analytically in \cite{new2}.

\subsection{Motion within the Family -- Asymptotic Matching}
\label{matching}
Above we stated that the acceleration of
holes in the perturbed CGLE
can be (approximately) described as a motion within the family of
Nozaki-Bekki hole solutions by taking the family parameter
$v=v(t)$ as slow variable.
This would, however, lead to
a change of the asymptotic wavenumbers
(one has $\dot  k = \hat k\dot v$)
which is not possible in an infinite system.
(As pointed out before the family mode
${\mit \Phi}_{fam} =\partial_v A_{v}(\zeta,t)$ ,
eqs.(\ref{fam_bed},\ref{phi_fam}),
describing the difference between two "neighboring" holes
diverges linearly for large $|\zeta|$ .)
The difficulty is resolved by limiting the
"motion within the family"
to a finite region of size $\sim \dot v^{-1}$
around the hole core (inner region).
A global solution for $A$ has
then to be constructed by asymptotic matching.
For calculating the acceleration, however, it will not be
necessary to perform the matching explicitly.

To illustrate this we consider the eigenvalue equation
(\ref{eigenvalue}) with the condition
\begin{eqnarray}
  \label{bound_global}
  |W| < \ \mbox{bounded} \quad \mbox{for}
    \quad \zeta\rightarrow\pm\infty
\end{eqnarray}
describing the acceleration of the hole
in the limit of small velocities (see next section).
In the outer region
\begin{eqnarray}
  \label{outer}
| \kappa \zeta | \gg 1
\qquad
  \mbox{(outer region)}
\end{eqnarray}
$W$ behaves exponentially with exponents  $p_i= p_i(\lambda)$
(c.f. Sec.\ref{lin_op}, App.\ref{app_pws}).
For values $0<\lambda\ll 1$
there are two decaying $(p_{1/2})$ and two growing $(p_{3/4})$
exponents so that boundary condition (\ref{bound_global})
is indeed equivalent to
\begin{eqnarray}
  \label{bound_inner}
  \mbox{
   '$W$ should not grow in space exponentially
    (with the exponents $p_{3,4}$)'}
\; \; .
\end{eqnarray}
It is crucial to note that because of the overlap of regions
the realization of this boundary condition
can already be controlled in the inner region without performing
any matching.

Since there is one weakly decaying exponent
$ p_1 \sim \lambda \sim \dot v $
boundary condition (\ref{bound_inner}) in practice leads
to an algebraic growth of $W$
(at the outer limit of the inner region).
This can be understood as a truncated
expansion of the exponentially decaying outer solution
(see eq.(\ref{a214}))
in the overlap region
$ \;
  W \sim
       \left( i + O(\lambda) \right) \;
       \left( 1+p_1\zeta + O(\lambda^2) \right) \; . \;
$
The portion $ \sim i p_1\zeta $
corresponds to the linear divergence of $\Phi_{fam}$.
{}From this one finds the outer limit of the inner region
\begin{eqnarray}
  \label{inner}
  |p_1 \zeta| \ll 1
  \qquad
  \mbox{(inner region)} \; .
\end{eqnarray}

Boundary condition (\ref{bound_inner}) is not restricted to
the case of small velocities.
Applied to the inner region it remains also
valid in more realistic situations
with slowly varying wavenumbers governed by phase equations
far away from the hole core.
We note that besides perturbations of the equation
also core instability or modified boundary conditions
(coming i.e. from the presence of a shock)
in general lead to a hole acceleration
which can be treated by an analogous matching approach.

\section{Perturbation Analysis}
\label{pert}

\subsection{Formal Analysis of Stationary Instabilities}
\label{stat_bif}
In this section we will investigate stationary instabilities
of the standing hole with respect to localized modes.
The formalism will be applicable to the acceleration instability
in the perturbed CGLE and to the core instability
in the cubic CGLE ($d=0$).

Let ${\bf u}= (b,c,d)$
denote a point in parameter space of the CGLE .
We write ${\bf u}= {\bf \hat u} + \Delta {\bf u}$
where ${\bf \hat u}$ belongs to the stability threshold.
In particular ${\bf \hat u}$ may lie on the core-instability line
$(b_{crit},c_{crit},d=0)$,
then $\Delta {\bf  u} = (\Delta b , \Delta c , 0 )$
is of interest (core instability),
or ${\bf \hat u}$ can be a point
$(b,c,0)$ where the standing hole is stable and then
$\Delta {\bf  u} = ( 0, 0 , d )$ (acceleration instability).
We want to construct the unstable localized mode
$W(x; {\bf u}= {\bf \hat u} + \Delta {\bf u})$
near the threshold.
It satisfies
\begin{eqnarray}
  \label{lin}
  L_0({\bf u})  W({\bf u})  =  \lambda({\bf u}) W({\bf u})
\end{eqnarray}
where $ L_0 $ has been introduced in eqs.(\ref{stab},\ref{L_eck}).

In the following we want to expand $W({\bf \hat u} + \Delta {\bf u})$
in powers of $\Delta {\bf u}$.
The calculations will be limited to the inner region (eq.\ref{inner})
where one may use boundary condition (\ref{bound_inner}) instead of
the usual condition (\ref{bound_global}) for localized modes.
We will use the symbol $\Delta {u}$ to denote the distance from
threshold along a fixed path which crosses the
neutral curve.
Consistently the derivative $\frac{\partial \,}{\partial u}$
signifies the derivative along this path.
At the threshold,
$\lambda({\bf \hat u})= 0$,
we have
\begin{eqnarray}
  \label{tresh}
  W({\bf \hat u}) = \Phi_{tr}({\bf \hat u})
  \; .
\end{eqnarray}
To see this we note that,
since $W({\bf \hat u}) $ belongs to
the null space of $L_0({\bf \hat u})$,
it has to coincide with one of the two
neutral modes given in eq.(\ref{neutral}).
For reasons of symmetry
one expects  to have $\Phi_{tr}$, which
destroys the zero at the hole core.

Inserting the expansion
\begin{eqnarray}
  W({\bf \hat u} + \Delta {\bf u}) =
  \Phi_{tr}({\bf \hat u} + \Delta {\bf u})
  + \Delta { u} W_1({\bf \hat u})
  + \Delta { u}^2 W_2 ({\bf \hat u})
  + O(\Delta { u}^3)
\end{eqnarray}
into eq.(\ref{lin}) one finds at first order in $\Delta { u}$
\begin{eqnarray}
  \label{o1}
  L_0 ({\bf \hat u})  W_1({\bf \hat u}) =
  \frac{ \partial\lambda({\bf\hat u})}
  {  \partial { u}} \; \Phi_{tr}({\bf \hat u})
  \,\, .
\end{eqnarray}
Equation (\ref{o1}) is proportional to eq.(\ref{fam_bed})
and thus has the solution
\begin{eqnarray}
  W_1 = \frac{ \partial\lambda({\bf u})}
  {\partial { u}} \Big|_{{\bf u}={\bf \hat u}} \Phi_{fam}
  ({\bf \hat u})
  \; .
  \label{w1}
\end{eqnarray}
At second order in  $\Delta {u}$ eq.(\ref{lin}) then yields:
\begin{eqnarray}
  \label{o2}
  L_0({\bf \hat u}) \left( W_2({\bf \hat u}) -
  \frac{1}{2}
  \frac{ \partial^{2}\lambda({\bf\hat u})}
  {  \partial { u}^{2}}  \;  \Phi_{fam}({\bf \hat u}) \right)
  =
  \frac{ \partial\lambda({\bf\hat u})}
  {\partial { u}}
  \left(
  \frac{ \partial\lambda({\bf\hat u})}
  {\partial { u}}
  \Phi_{fam}({\bf\hat u})
  +
  \left[ \frac{ \partial\Phi_{tr}({\bf\hat u})}
  {  \partial { u}}
  -
  \frac{ \partial L_0({\bf\hat u})}
  {  \partial { u}}
  \Phi_{fam}({\bf\hat u})
  \right]
  \right)
\end{eqnarray}
where the term
$\frac{1}{2}\frac{ \partial^{2}\lambda({\bf\hat u})}
{\partial {u}^{2}}
\Phi_{tr}({\bf\hat u})$
was brought to the lhs with the help of eqs.(\ref{o1}) and (\ref{w1}).
If a function
$\Phi_{fam}({\bf \hat u}+ \Delta {\bf u})$ with
\begin{eqnarray}
  \label{fam}
  L_0 ({\bf \hat u}+ \Delta
  {\bf u})  \Phi_{fam}({\bf \hat u}+ \Delta
  {\bf u}) =
   \Phi_{tr}({\bf \hat u} + \Delta
  {\bf u}) \;.
\end{eqnarray}
exists at ${\bf \hat u}+ \Delta{\bf u}$
the term in square brackets can be absorbed
by modifying  $W_2$.
To see this one may
expand $W$ in the following way
\begin{eqnarray}
  W({\bf \hat u} + \Delta {\bf u}) =
  \Phi_{tr}({\bf \hat u} + \Delta
  {\bf u})
  + \Delta { u}  \frac{ \partial\lambda({\bf\hat u})}
  {\partial { u}}
  \Phi_{fam}({\bf \hat u}+ \Delta
  {\bf u})
  + \Delta { u}^2 {\tilde W}_2 ({\bf \hat u})
  + O(\Delta { u}^3)
\end{eqnarray}
which yields at second order in $\Delta u$ :
\begin{eqnarray}
  \label{o2b}
  L_0({\bf \hat u}) \left(\tilde W_2({\bf \hat u})
  - \frac{1}{2} \frac{ \partial^{2}\lambda({\bf\hat u})}
  {  \partial { u}^{2}}  \;  \Phi_{fam}({\bf \hat u}) \right)
   =
  \left(\frac{ \partial\lambda({\bf\hat u})}
  {\partial { u}}
  \right)^2
  \Phi_{fam} ({\bf \hat u})
\end{eqnarray}
Equation (\ref{fam}), which is the condition for the existence
of a hole family at ${\bf \hat u}+ \Delta{\bf u}$,
is satisfied near the core-instability line if $d$
is kept zero.
Thus the solvability of eq.(\ref{o2b}) gives a criterion
for the
threshold of the core instability ($ d=0$)
which will be evaluated fully analytically in the next section.
Near the threshold of the acceleration instability
however eq.(\ref{o2}) can only be fulfilled if
\begin{eqnarray}
  \frac{ \partial\lambda({\bf\hat u})}
  {\partial { u}}
  =
  \frac{ \partial}
  {\partial {d}}
  \left[\frac{\dot v}{v}
  \right]_{d=0}
\end{eqnarray}
is adjusted properly which can be used to determine
the acceleration of the holes.
Thus the quantitative results for the acceleration of the holes
given  below represent a proof for the destruction of the hole family
by the higher-order perturbation.

{}From eq.(\ref{o2}) one also sees that $\dot v/v$
diverges as one approaches the core-instability line in the $b$-$c$
plane. Near this line however the analysis of this chapter
looses its validity. Here the two modes leading to core and
acceleration instability interact and the scenario becomes
more complex (see eq.(\ref{phen})).

\subsection{Core Instability}
\label{core_inst}
In this section we calculate the core instability line in the cubic CGLE.
We do this by giving a solvability condition for
eq.(\ref{o2b}) which we write in the form
\begin{eqnarray}
  \label{solv_core_1a}
  L({\bf \hat u}) W({\bf \hat u})  = \Phi_{fam}({\bf \hat u})
\; .
\end{eqnarray}
Here $ \Phi_{fam} $ is the family mode from eq.(\ref{phi_fam})
and $ L({\bf \hat u}) = L^{NB}_0 (b,c) $ is the linear operator around
a standing hole (see eq.(\ref{Lv})).
Equation (\ref{solv_core_1a}) reads explicitly
\begin{eqnarray}
  \label{solv_core_1}
  (1 \! + \! ib) \; \partial_{\zeta \zeta} W
  \; + \;
  (2 i  K (1 \! + \! ib) \tanh (\kappa \zeta)) \; \partial_{\zeta} W
  \; +  \;
  (1 + i \Omega + (1 \! + \! ib) i  K \kappa )
  \frac{1}{\cosh^2(\kappa \zeta)} \;  W
  \; - \;
  \qquad \qquad \qquad \nonumber    \\
  \qquad \qquad  \qquad
  - \; (1 \! + \! ic) (1 \! - \!  K^2) \tanh^2 (\kappa \zeta) \;
  (W+W^*) \; \; = \; \;
  -  K - 2 \kappa i  \; + \;
  i (1 \! - \!  K^2) \zeta \tanh (\kappa \zeta)
\end{eqnarray}
Solvability here  means that we have to adjust
the parameters $ b = b(c) $ in a way that the
(symmetric) solution $ W $ of this equation
satisfies boundary condition (\ref{bound_inner}).
In a first step we simplify eq.(\ref{solv_core_1}) by making the
following ansatz for $ W $ ("variation of constants")
\begin{eqnarray}
  \label{var_const}
  & W  = f_{rot} \Phi_{rot} + f_{tr} \Phi_{tr} &
  \nonumber   \\
  & \qquad \mbox{where} \quad
  \qquad \qquad
  \Phi_{rot} = i \tanh (\kappa \zeta),
  \qquad
  \Phi_{tr} = 1/\cosh^2 (\kappa \zeta) \, +
                 \, (i/{\hat \kappa}) \, \tanh^2 (\kappa \zeta)  &
\end{eqnarray}
are the two neutral modes of $ L_{0} $
(c.f. eqs.(\ref{phi_rot},\ref{phi_tr})).
$ f_{rot} $ and $ f_{tr} $
are two real functions to be determined.
Inserting this ansatz into eq.(\ref{solv_core_1}) one gets
after elimination of $ f_{rot} $ the following real
second-order differential equation for $ f := \partial_{\zeta} f_{tr} $
\begin{eqnarray}
  \label{solv_core_2}
  & \hat L f = I_{fam} &
    \nonumber \\
  & \mbox{where} \quad
    \hat L \; := \;
    \partial_{\zeta \zeta}
    - 6 \kappa \tanh (\kappa \zeta) \partial_{\zeta}
    + \left( (6  K^2 + 12 \kappa^2) \tanh^2 (\kappa \zeta)
    - 4 \kappa^2 \right) & \nonumber \\
  & \qquad I_{fam} \; := \; 6  K^2 \zeta \cosh^2(\kappa \zeta)
    + 3  K^2 (b \hat \kappa - 2) \zeta
    + (b + b  K^2 - 4  K^2 \hat \kappa) 3 K/(1- K^2) \; \tanh(\kappa \zeta)
      \cosh^2(\kappa \zeta) &
\end{eqnarray}
{}From the asymptotic behavior ($|\zeta| \to \infty$)
of eq.(\ref{solv_core_2}) one easily sees that
the only boundary condition for the (antisymmetric)
function $ f = \partial_{\zeta}(\cosh^{2}(\kappa\zeta)Re(W))$
which is compatible with condition (\ref{bound_inner}) for $W$
is given by
\begin{eqnarray}
  \label{bound_f}
  f \sim \zeta e^{2 \kappa |\zeta|}
  \quad \mbox{for} \; \zeta \rightarrow \pm \infty
\; .
\end{eqnarray}
This excludes the behavior
$ \; \sim e^{(2\kappa + p_{3/4}) |\zeta|} $
of the two fundamental modes of $ \hat L $.
Equations (\ref{solv_core_2},\ref{bound_f})
can be solved with the ansatz
\begin{eqnarray}
  \label{core_ansatz}
  f & = & \sum_{m=-1}^{\infty} c_m \zeta \cosh^{-2m}(\kappa \zeta) \; + \;
          \sum_{m=-1}^{\infty} d_m \tanh(\kappa \zeta) \cosh^{-2m}(\kappa
\zeta)
\end{eqnarray}
As shown in Appendix \ref{app_core} for certain values
of $(b,c)$ with $b=b(c)$
one can adjust the coefficients $ \; c_m, \; d_m $
in a way that the series converges uniformly towards a
smooth function solving the boundary value problem
eqs.(\ref{solv_core_2},\ref{bound_f}).
The condition for convergence defines the core instability line.
Explicitly this line is given by
\begin{eqnarray}
  \label{core_line}
  & \frac{1 \! - \!  K^2}{1+\hat\kappa^2} \; S(\hat\kappa^2) \; + \;
  1 \; + \;  K^2 \frac{3b - 14 \hat\kappa}{3b + 2 \hat\kappa}
  \; = \; 0
  &
\end{eqnarray}
\begin{eqnarray}
  \mbox{where} \quad
    S (\hat\kappa^2)  & := &
    \frac{3}{4+3/\hat\kappa^2} \; +
    \;\sum\limits_{n=2}^{\infty} \frac{2n+1}{2n^2+2n+3/\hat\kappa^2} \;
    \prod\limits_{m=2}^n \frac{2m^2-m+3/\hat\kappa^2}{2m^2+m+3/\hat\kappa^2}
              \\
  & = &
    \frac{3}{4+3/\hat\kappa^2} \; +
    \; \frac{\Gamma(\frac{9}{4}+a) \Gamma(\frac{9}{4}-a)}
            {\Gamma(\frac{7}{4}+a) \Gamma(\frac{7}{4}-a)} \;
    \;\sum\limits_{n=2}^{\infty} \frac{2n+1}{2n^2+2n+3/\hat\kappa^2} \;
       \frac{\Gamma(n+\frac{3}{4}+a) \Gamma(n+\frac{3}{4}-a)}
            {\Gamma(n+\frac{5}{4}+a) \Gamma(n+\frac{5}{4}-a)}
    \nonumber \\
  &   &
    \mbox{with} \;\; a:= \frac{1}{4}\sqrt{1-24/\hat\kappa^2}
    \nonumber
\end{eqnarray}
In the limit of large $c$ one can neglect the first term
in eq.(\ref{core_line})
which contains the infinite sum $S(\hat\kappa^2)$
and eq.(\ref{core_line}) reduces to
\begin{eqnarray}
  \label{core_lim}
  \mbox{either} \qquad & &
    b^4 = \frac{16}{9} c^2 + O(c^{3/2})  \quad
   \mbox{for} \; \; c \rightarrow  \infty
   \quad (b>0)
   \\
  \mbox{or} \qquad & &
    b = - 1/\sqrt{2} + O(1/c) \quad
   \mbox{for} \; \; c \rightarrow  \infty
\end{eqnarray}
Eqs.(\ref{core_lim}) describe both branches of the core instability line
in the limit $ c\rightarrow \infty $.
The core instability line (\ref{core_line}) together with the
curve describing the instability of the plane waves at the wings of
the hole give the complete stability diagram of the standing
Nozaki-Bekki hole solution. The curves are plotted in Fig.1.
They are consistent with previous results
\cite{S85} \cite{MC} \cite{sasa}
and generalize them.
Finally we remind that our derivation made use of
the fact that the core instability occurs via a static bifurcation,
which was first found by \cite{MC} and \cite{sasa}.

\subsection{Acceleration Instability}
\label{acc_inst}
In the following we treat the acceleration of a moving hole
caused by a perturbation of the cubic CGLE.
We use adiabatic elimination which
in the limit $v\to 0$
can be put into the framework introduced in section \ref{stat_bif}.
We assume that in the perturbed CGLE the hole acceleration occurs
near the hole family and can be described by taking
$ v=v(t) $ as the slow variable while the other degrees
of freedom follow adiabatically.
Formally this is done by writing
\begin{eqnarray}
  \label{acc_inst_1}
  A = ( Z_{v(t)}(\zeta) + W(\zeta) ) \;
      \exp \left(i \chi_{v(t)}(\zeta)
      - i \int_0^t \left(\Omega_{v(\tau)} + \Delta \! \Omega \right)
        d\tau \right)
  ; \qquad
  \zeta := x-\int_0^t v(\tau) d\tau
\end{eqnarray}
where $ Z_v, \chi_v, \Omega_v $ describe the slowly time dependent NB hole.
$ W , \Delta \! \Omega $ are the small changes of the solution caused
by the quintic perturbation.
At lowest order they are time independent
(adiabatic elimination).
As before this ansatz is only valid in the inner region
(\ref{inner}).
Inserting (\ref{acc_inst_1}) into the CGLE
(\ref{eq1frame}) in the co-moving frame rotating
with frequency $ \Omega_{v(t)} + \Delta \! \Omega $
one obtains
\begin{eqnarray}
  \label{acc_inst_2}
  \dot v \partial_{v} A
      & = &
      F_{v(t) \; \Omega_{v(t)} + \Delta \! \Omega} [A]
\end{eqnarray}
The only time dependence which cannot be transformed away
by this choice of the coordinate system is given by a movement
within the family $ \dot v \partial_{v} A $.
$ W, \dot v, \Delta \! \Omega $ are
of order $ d $.
Using the fact that
$ A^ {NB}_{v} $ solves the cubic CGLE
$  F_{v \; \Omega}^{d = 0} [A^ {NB}_{v}] = 0 $
at each time $ t $ and neglecting all higher order terms in
$d$
one arrives at the following ordinary linear
differential equation for the perturbation  $ W $
\begin{eqnarray}
  \label{acc_inst_3}
   L^{NB}_v W & = &  I_v(\Delta \! \Omega,\dot v)
    \nonumber \\
  \mbox{where} \qquad \qquad
      I_v(\Delta \! \Omega,\dot v) &:= &
      \dot v \Phi_{fam}
      - i \Delta \! \Omega Z_{v} - d |Z_{v}|^4 Z_{v}
\end{eqnarray}
Here one has to adjust the parameters
$ \Delta \! \Omega, \dot v $ in the inhomogeneity
in such a way that the solution $ W $
satisfies (\ref{bound_inner}) which
determines the acceleration $\dot v$.

Equations (\ref{o2}) and (\ref{acc_inst_3}) can both serve to
calculate the acceleration
sufficiently far away from the core instability line
(where it is possible to replace eq.(\ref{phen})
 by eq.(\ref{phen_a})).
While eq.(\ref{acc_inst_3}) is applicable for holes moving
with arbitrary velocities $v$ (but small $\dot v$)
eq.(\ref{o2}) is only valid for small $v$.
The connection between both equations can be seen
by expanding (\ref{acc_inst_3}) in terms of $v$.
Writing
\begin{eqnarray}
  \label{acc_inst_4}
   L^{NB}_v & = & L^{NB(0)}_v + v L^{NB(1)}_v + O(v^2)
      \nonumber \\
   W        & = & W^{(1)} + v W^{(2)} + O(v^2)
      \nonumber \\
   I_v      & = & I^{(1)} + v I^{(2)} + O(v^2)
\end{eqnarray}
and remembering that $L^{NB(0)}_v=L^{NB}_0$ is the
linear operator from eq.(\ref{solv_core_1}) describing
perturbations around the standing hole of the cubic CGLE,
eq.(\ref{acc_inst_3}) becomes at order $v^0$
\begin{eqnarray}
  \label{acc_inst_5}
  L^{NB}_0 W^{(1)} & = & I^{(1)}
     \nonumber \\
  & = & - i \Delta \! \Omega Z_0 - d Z_0^5
\end{eqnarray}
Here $I^{(1)}$ and thus $ W^{(1)}$ are antisymmetric.
$W^{(1)}$ has to satisfy the boundary condition
(\ref{bound_inner}) which fixes the parameter $\Delta \! \Omega$.
$W^{(1)}$ and $\Delta\! \Omega$ describe the changes of a
standing hole resulting from
a $ d $--perturbation in the cubic CGLE.
These changes have already been included in the approach
of subsection \ref{stat_bif}.
$W^{(1)}$ and $\Delta\! \Omega$ can be found analytically
(see eq.(\ref{acc_nls_13})).
At order $ v^1 $ eq.(\ref{acc_inst_3}) then reads
\begin{eqnarray}
  \label{acc_inst_6}
  L^{NB}_0 W^{(2)} & = & - L^{NB(1)}_v W^{(1)} + I^{(2)}
        \nonumber \\
  & = & - L^{NB(1)}_v W^{(1)}
        + \frac{\dot v}{v} \Phi_{fam}
        - i \Delta \! \Omega \hat A
        - d Z_0^4 (3 \hat A + 2 \hat A^*)
\end{eqnarray}
This equation corresponds to eq.(\ref{o2}).
The r.h.s., and thereby the
perturbation $W^{(2)}$, are symmetric
and one has to adjust the acceleration $ \frac{\dot v}{v} $
such that $ W^{(2)} $ fulfills the boundary condition
(\ref{bound_inner}).
(Note that in deriving eq.(\ref{acc_inst_6}) we have expanded
the equation of motion (\ref{acc_inst_2}) in $d$ and
afterwards in $v$ whereas eq.(\ref{o2}) can be found from
eq.(\ref{acc_inst_2}) by an interchange of the expansions.)
For small $ d , v $ one has
\begin{eqnarray}
  \label{acc_inst_7}
  \frac{\dot v}{v} = Re(g^* d) + O(v^2)
  \qquad g:= g_1 + i g_2 \qquad (g_i \; \mbox{real})
\end{eqnarray}
In Section \ref{num_meth} eq.(\ref{acc_inst_3}) will be used to determine
$ g $ numerically for arbitrary $ b,c $ while in App.\ref{nls_acc_inst}
eq.(\ref{acc_inst_6}) will be used to calculate $ g $ analytically
in the NLS-limit $ ( b,c \rightarrow \infty) $.

\subsection{Hole -- Shock Interaction}
\label{int_hs}
The acceleration of a hole as calculated in the last
subsection is influenced
in the presence of a shock.
Here we treat this acceleration in
the limit of large hole--shock separation.
For that purpose we divide space into
overlapping hole and shock regions.
The important point is that when using the ansatz (\ref{acc_inst_1})
in the hole region the boundary condition (\ref{bound_inner})
is changed by the shock.
In the region of overlap one now has
(besides decaying $ \; \sim  e^{p_{1/2} \zeta} $)
growing stationary perturbations
$ \; \sim  e^{p_{3/4} \zeta} $
forming the shock out of the plane-wave state.
The distance from the shock
determines the \underline{magnitude} of the growing perturbation
(prefactor of $W$)
and thus the hole acceleration.

The hole--shock interaction depends crucially on the fact
whether the growing exponents $p_{3/4}$ are real or complex conjugated.
Accordingly one has monotonic ('monotonic case')
or oscillatory ('oscillatory case') interaction
(see below eqs.(\ref{int_4},\ref{int_5})).
The boundary between the two interaction types in the
$b,c$--parameter plane is given by the condition
$p_3=p_4$
which depends on the hole velocity $v$ since
$p_{3/4}=p_{3/4}(v)$ (c.f.App.\ref{app_pws}).
In particular for a standing hole the boundary line
becomes
\begin{eqnarray}
   \label{int_1}
   \hat \kappa^2 - 6 = 0
\end{eqnarray}
which is included in Fig.1
($\hat \kappa$ is given in eq.(\ref{kappa}).).

In the overlap range ($\kappa^{-1} \ll \zeta \ll L$)
between the hole position $\zeta = 0$ and the
shock position $\zeta=L$
(defined by $\partial_{\zeta}arg(A)|_{\zeta=L} = 0$)
one has approximately plane waves plus
linear perturbations $W$ of the form
(c.f.App.\ref{app_pws})
\begin{eqnarray}
   \label{int_2}
   W e^{-i\phi} & = & r_3 e^{p_{3}\zeta + i\phi_3} +
           r_4 e^{p_{4}\zeta + i\phi_4}
   \qquad
   \mbox{(monotonic case)}
   \\
   \label{int_3}
   \mbox{or} \quad
   W e^{-i\phi} & = & z e^{p_{3} \zeta} +
           \eta(p_{3}) z^* e^{p_{3}^* \zeta}
   \qquad
   \mbox{(oscillatory case)} .
\end{eqnarray}
(Here $\phi:=arg(v\hat A +  K \hat B)$ is a fixed phase factor.)
The perturbations are made up of the two growing fundamental
modes of $L_v$.
The two real prefactors $r_{3/4}=r_{3/4}(L)$
or the complex constant $z=z(L)$
describing the contribution of these modes
have to be determined from the asymptotic shape of the
shock solution (see below).
They depend exponentially
on the distance $L$ between hole and shock
with reversed exponents $\;\sim e^{-p_{3/4}L}$.
Thus the $L$ dependence of the acceleration induced
by the perturbation $W$ is given by
\begin{eqnarray}
   \label{int_4}
   \dot v & = & g_3  e^{-p_{3}L} + g_4  e^{-p_{4}L}
   \qquad
   \mbox{(monotonic case)}
   \\
   \label{int_5}
   \mbox{or} \quad
   \dot v & = & g_3 e^{-Re(p_{3})L} \sin(Im(p_{3})L + g_4)
   \qquad
   \mbox{(oscillatory case)}
   \quad .
\end{eqnarray}
The two unknown real constants $g_3,g_4$ have to be determined
from solving eq.(\ref{acc_inst_3})
with boundary condition (\ref{int_4},\ref{int_5})
instead of (\ref{bound_inner}).
In the monotonic case the smaller
exponent $p_4$ ($p_3>p_4>0$) dominates asymptotically and
for practical purposes the coefficient
$g_4$ is (usually) sufficient to describe the acceleration.

Now we turn to the calculation of the constants
$r_{3/4}$ or $z$ from the asymptotics of the shock solution.
Besides the long--wavelength approximation,
where after a Hopf Cole transformation
the shock region is described by a linear phase equation
(see \cite{new2}),
there exists no analytic expression for the shock solution.
Therefore in general the constants $ r_3,r_4,z $ have to be
calculated numerically by solving a boundary value problem
(c.f.Sec.\ref{num_meth}).
In addition a crude analytic estimate can be obtained by the
following 'linear approximation':
we extend the hole region up to the
center of the shock which means that the shock is
a superposition of a plane wave
and the linear perturbations $ \sim e^{p_{3/4}\zeta} $.
Assuming $\partial_{\zeta} A|_{\zeta=L}=0$
at the position $ \zeta = L $ of the shock
(which corresponds to a standing shock)
one gets by using again (\ref{acc_inst_1})
\begin{eqnarray}
   \label{int_6}
   \partial_{\zeta} \left.
       \left( (Z_{v(t)} + W) e^{ i \chi_{v(t)}} \right)
       \right|_{\zeta = L} & = & 0
       \nonumber  \\
   \stackrel{\kappa L \gg 1}{\Longleftrightarrow} \quad
       \left. \left( \partial_{\zeta} W
       + i q_1 \left(v \hat A +  K \hat B + W \right) \right)
       \right|_{\zeta = L} & = & 0
\end{eqnarray}
Inserting $ W $ from eq.(\ref{int_2}) or (\ref{int_3})
one obtains an approximate expression for
$ r_{3/4}(L) $ or $ z(L) $ respectively.

So far we have only treated the interaction of
one hole and one shock in the cubic CGLE.
In general boundary conditions
(\ref{int_2},\ref{int_3}) apply
on both sides of the hole separately with
corresponding shock distances $ L_{(r/l)} $.
In most cases, however,
one of the hole--shock interactions dominates
and the constants $r_{3/4}$ or $z$ can be set zero
on the other side
(corresponding to the interaction with
 a shock infinitely far away).
In any case the full information on the boundary
condition for $ W $ can be gathered into one real
four vector $\vec b$
containing the boundary constants
$ r_3,r_4 $ or $ z $ for both sides.
Since in the perturbed CGLE the
hole acceleration depends linearly on this
boundary vector $\vec b$ and on the
$d$--perturbation
(via the inhomogeneity of eqs.(\ref{acc_inst_3},\ref{acc_inst_6}))
one may simply add the hole accelerations
caused by neighboring shocks (eqs.(\ref{int_4},\ref{int_5})) and
$d$--perturbations of the CGLE (eq.(\ref{acc_inst_7})).
This is exploited in the next section.

\section{Numerical Results and Comparison with Simulations}
\label{comparison}
\subsection{Numerical Methods}
\label{num_meth}
We now show how one can determine numerically the hole
acceleration induced by a $ d $-perturbation or an
interaction with a shock.
Mathematically this means solving the differential
equation (\ref{acc_inst_3}) with the
boundary conditions (\ref{bound_inner})
or (\ref{int_2},\ref{int_3}) described by the
boundary vector $ \vec b $.

First let us assume the vector $ \vec b $ were known.
One has to adjust $ \dot v , \Delta\!\Omega $ in eq.(\ref{acc_inst_3})
in a way that the solution $ W $ is asymptotically described
by $ \vec b $.
This can be achieved by the following 'shooting' method.
One integrates eq.(\ref{acc_inst_3}) (numerically)
starting at $ \zeta =0 $.
The correct initial values $ W(0) $ and
$ \partial_{\zeta} W (0) $ are unknown but one may assume without
loss of generality
$ Re(W(0)) = 0 $ and $ Im(\partial_{\zeta} W(0)) = 0 $,
which can always be achieved by a suitable addition
of the neutral modes $ \Phi_{rot} $ and $ \Phi_{tr} $
to the perturbation $ W $.
One is left with two real parameters
$ Im(W(0)) $ and $ Re(\partial_{\zeta} W(0)) $ describing
the initial condition and two further real parameters
$ \dot v , \Delta\!\Omega $ occuring in
the inhomogeneity of eq.(\ref{acc_inst_3}).
Now one integrates eq.(\ref{acc_inst_3}) from $ \zeta =0 $ once to
(sufficiently large) positive $\zeta$ and once to negative $\zeta$
with all four parameters set zero.
{}From the asymptotic growth of $ W $ at the wings one can
read off a real four vector $ \vec b_0 $ describing this growth.
Doing similar integrations with the following values of the parameters
one obtains the boundary vectors $ \vec b_i ,\; (i=1...4) $.
\begin{eqnarray}
  \label{mm_1}
  \begin{array}{cccccc}
    Im(W(0))=1 \quad  &  Re(\partial_{\zeta} W(0))=0 \quad  &  \dot v =0 \quad
&
       \Delta\!\Omega=0 \quad  &  \Rightarrow \quad  &  \vec b_1  \\
    Im(W(0))=0 \quad  &  Re(\partial_{\zeta} W(0))=1 \quad  &  \dot v =0 \quad
&
       \Delta\!\Omega=0 \quad  &  \Rightarrow \quad  &  \vec b_2  \\
    Im(W(0))=0 \quad  &  Re(\partial_{\zeta} W(0))=0 \quad  &  \dot v =1 \quad
&
       \Delta\!\Omega=0 \quad  &  \Rightarrow \quad  &  \vec b_3  \\
    Im(W(0))=0 \quad  &  Re(\partial_{\zeta} W(0))=0 \quad  &  \dot v =0 \quad
&
       \Delta\!\Omega=1 \quad  &  \Rightarrow \quad  &  \vec b_4
  \end{array}
\end{eqnarray}
Because of the linearity of the boundary value problem one can now
determine the set of parameters corresponding to the desired
boundary vector $ \vec b $.
Solving the linear equation
\begin{eqnarray}
  \label{mm_2}
  \vec b_0 +
  Im(W(0))                  (\vec b_1 - \vec b_0) +
  Re(\partial_{\zeta} W(0)) (\vec b_2 - \vec b_0) +
  \dot v                (\vec b_3 - \vec b_0) +
  \Delta\!\Omega            (\vec b_4 - \vec b_0) =
  \vec b
\end{eqnarray}
yields the acceleration $ \dot v $ and the frequency
shift $ \Delta\!\Omega $ , which the hole receives under the influence
of neighboring shocks and/or a $ d $-perturbation.
One needs two of these runs for the calculation of $ g_1,g_2 $
(eq.(\ref{acc_inst_7})) and two further runs for the calculation of
$ g_3,g_4 $ (eqs.(\ref{int_4},\ref{int_5})).

Now we turn to the determination of the constants $ r_3,r_4,z $
in eqs.(\ref{int_2},\ref{int_3}).
Since the shock solution connects two plane--wave states
it is described by the ODE (\ref{eq1frame})
$ F_{v_s \Omega_s}^{d=0} [A] = 0 $ with the shock
velocity $ v_s $ and
frequency $ \Omega_s $ fixed by the (given) wavenumbers
$ q_1,q_2 $ at the wings.
We solved this boundary value problem using a NAG routine.
In some intermediate region between the system boundaries and the
center of the shock
the amplitude is just given by the incoming plane waves
and linear perturbations $ \sim e^{p_{3/4} \zeta} $ growing
towards the shock center.
In this region we extract the desired constants $ r_3,r_4,z $.

Our simulations of the CGLE (\ref{eq1}) were performed with a pseudospectral
code based on FFT using a predictor corrector scheme in time.
We had to use very high precision in the simulations, because
discretization errors, like the $ d $-perturbation,
in general destroy the inner
symmetry of the cubic CGLE and therefore have similar effects on the
hole solution, i.e. they also lead to an acceleration.
Typically we used discretizations of $ \Delta x \sim 0.05-0.15 $ and
$ \Delta t \sim 0.001-0.02 $ depending on the parameters $ b,c,d $
of the CGLE.

\subsection{Results}
\label{results}
Next we compare the results from the semianalytical
calculations in sections \ref{acc_inst},\ref{int_hs}
with those of direct simulations of the CGLE.

Figure 2 shows the acceleration of a hole center caused by a quintic
perturbation of the CGLE.
The acceleration formula (\ref{acc_inst_7}) with the constant
$g$ obtained from the numerical matching
(using eqs.(\ref{acc_inst_3},\ref{bound_inner}))
is seen to be in good agreement with the simulations.
This also confirms that the acceleration instability
is indeed of stationary type.
When scanning the $ (b,c) $--range of stable hole solutions
one always finds deceleration for (real) $ d < 0 $,
consistent with the simulations.
{}From the fully analytical treatment in App.\ref{nls_acc_inst}
(c.f. eq.(\ref{acc_nls_14})) it is seen that this result remains
valid in the nonlinear Schroedinger limit.
(Then one has numerical problems with the matching
 because here the ratio $ |p_3/p_4| $ of the two growing
 exponents $ p_3, p_4 $ tends to infinity which prevents
 a correct numerical extraction of the boundary vectors $b_i$
 (c.f. section \ref{num_meth}).)

As already stated in section \ref{int_hs}
the $ (b,c,d) $--region of stable hole solutions is divided
into parts with monotonic
and with oscillatory interaction between holes and shocks.
Figures 3a,3b show the acceleration of a standing hole
under the influence of a standing shock solution for both
interaction types in the case $d =0$.
Here in the direct simulations the shock solutions were simply modeled
by taking $ \partial_{\zeta} A = 0 $ at the system boundaries.
As a result one has almost perfect quantitative agreement of simulations and
numerical matching if in the latter the shock is treated nonlinearly,
i.e. when the parameters $ r_3,r_4 $ (or $z$) in the boundary
conditions (\ref{int_2},\ref{int_3}) are extracted from a
numerical solution of the shock boundary value problem.
With the 'linear approximation' (\ref{int_6}) one obtains only
qualitative agreement.
When scanning the monotonic interaction range with the matching routine
we found that the asymptotic hole shock interaction is always attractive.
Near the nonlinear Schroedinger limit one has monotonic
interaction (see Fig.1).
Here one can calculate analytically
the hole-shock interaction
(see \cite{new2}) and it also turns out to be always attractive.

\section{Arrangements of Holes and Shocks}
\label{per_sol}
In this section we use the results from the previous sections for
the description of stable states made up of a periodic arrangements
of holes and shocks.
Such states
are frequently observed in
simulations with periodic boundary conditions
(see e.g.\cite{chate},\cite{PRL}).
Figure 4 shows the modulus $|A|=|A(x)|$ of a typical solution
found in a simulation.
As shown in Fig.5 one finds uniform as well as (almost)
harmonic and strongly anharmonic oscillating hole velocities.
(Slightly) beyond the core instability line the direction of the
velocity is changed in the oscillations (see Fig.5d).
The solutions are seen to be very sensitive to
$d$--perturbations of the cubic CGLE.

The uniformly moving solutions can be
well understood from the results of the last sections.
First one has to note that they are not expected
to be stable (and we indeed could not observe them in simulations)
in the monotonic range
\footnote{The boundary between monotonic and oscillatory interaction
          depends on the hole velocity (see \ref{int_hs}).}
since here the asymptotic hole--shock interaction is always attractive
(c.f. section \ref{results}).
Therefore we may restrict ourselves to the case of
oscillatory interaction.
Away from the core instability line uniformly moving
periodically modulated solutions can then
be identified as fixed points of a first order
differential equation for the hole velocity $v$
as the only slow variable.
The other degrees of freedom are thereby assumed to follow
adiabatically.

One equation for the hole velocity
$v$ is obtained by considering the
acceleration caused by interaction with
a neighboring shock at distance $L$
together with that resulting from
a $d$--perturbation.
{}From eqs.(\ref{acc_inst_7},\ref{int_5})
one obtains in the limit of small velocities $v$
\begin{eqnarray}
  \label{per_sol_2}
  \dot v \; = \; v \; Re(g^* d)
         \; + \; g_3 e^{-p_{3}'(v)L} \sin(p_{3}''(v)L + g_4) \; ,
         \qquad \quad   g=g_1+ig_2 \; ,
         \qquad         p_3=p_3'+ip_3''
\end{eqnarray}
Here we have taken into account only the interaction of the hole
with one of the neighboring shocks, implicitly assuming
that the period $P$ of the solution satisfies $P\gg L$
in addition to $\kappa L \gg 1$ , which is necessary
for the asymptotic analysis.
Equation (\ref{per_sol_2}) is correct up to first order in $v$ if
the two matching parameters $g_{3,4}$ are calculated including corrections
linear in $v$ and $\dot L $ in the matching procedure
(i.e. $ g_3(v,\dot L) = g_{30} + v g_{3v} + \dot L g_{3\dot L} $).
In the following we eliminate the two unknowns $L,\dot L$.

The shock velocity $v_s=v+\dot L$
is determined by the incoming
(plane) waves via a generalization of eq.(\ref{topol})
to slowly varying solutions leading to
\begin{eqnarray}
  \label{per_sol_3}
  v_s = v +  O(\dot v) \; .
\end{eqnarray}
Thus $\dot L = O(\dot v)$ can be neglected in the matching
parameters $g_3,g_4$ occurring in eq.(\ref{per_sol_2})
(since the factor $e^{-p_3'(v)L}$ is small).


{}From periodicity one has the condition that the phase
along one period $P$ of the solution
is an integral multiple of $2\pi$.
Using the fact that the full phase $\phi$
of a NB--hole solution is given by
\begin{eqnarray}
  \label{per_sol_4}
  \phi(\zeta) & = &
     \chi_v(\zeta) + arg(Z_v(\zeta)) \nonumber \\
  & \approx &
      K |\zeta| - \ln 2 +  k \zeta +
        arg(v\hat A \pm  K\hat B)
     \qquad \mbox{for} \;
     \pm \kappa \zeta \gg 1
\end{eqnarray}
one can derive the phase condition
\begin{eqnarray}
  \label{per_sol_5}
  2 \pi n
  \; \stackrel{!}{=} \;
     \phi(P)-\phi(0)
  \; \approx \;
     2  K L + P( k- K) +
     arg \left( \frac{v\hat A+ K\hat B}
                     {v\hat A- K\hat B} \right) \; + \;
     \phi_s(v)
  \qquad  (n \in {\bf N})
\end{eqnarray}
Here $\phi_s$ is the phase change due to the presence
of the shock.
For symmetry reasons $\phi_s=\phi_s(q_1,q_2)$
with $q_i=q_i(v)$
is an antisymmetric function of the hole velocity $v$
and it becomes linear for small velocities.
Solving (\ref{per_sol_5}) for $L$ and linearizing in $v$
one gets
\begin{eqnarray}
  \label{per_sol_6}
  L \approx L_{nP} - g_5 v
  \qquad \quad \mbox{with:} \; \;
  L_{nP} :=
   \frac{1}{2} \left( P + (2n-1)\pi/ K \right) ,
  \quad
  g_5 :=
   \frac{1}{2 K} \left(
   \left. \frac{\partial \phi_s}{\partial v} \right |_{v=0}
   + \frac{ k}{v}
     \left(P + \frac{2 \hat A''}{ K \hat B}\right) \right)
\end{eqnarray}
Inserting eqs.(\ref{per_sol_3}) and (\ref{per_sol_6}) into
eq.(\ref{per_sol_2}) one arrives at a differential
equation for $v$ of the desired form
\begin{eqnarray}
  \label{per_sol_7}
  \dot v = f_{nP}(v)
\end{eqnarray}
describing solutions with
$n,P$ as parameters in addition to $b,c,d$.
In the derivation we have assumed stationary
(or at most slowly evolving) solutions
as well as small velocities.
Furthermore we have neglected
$d$--corrections in the
plane wave quantities ($ k, K,p_{3,4},...$).
All parameters ($g_1,..,g_5$) can be extracted from
the matching procedure and the numerical nonlinear
shock solution.


The implications of eq.(\ref{per_sol_7}) are very simple.
The fixed points ($\dot v = 0$) describe the uniformly
traveling states.
Their velocity $v_{fp}=v_{fp}(n,P)$
and hole--shock separation $L_{fp}=L_{fp}(n,P)$
are predicted by eq.(\ref{per_sol_7})
as functions of $n,P$.
Their stability (within the family parameter $v$) is determined
by the sign of the linear growth rate
$\lambda_{fp} := \partial f_{nP}/ \partial v |_{v_{fp}}$.
In Figs.6,7 these theoretical predictions for
$v_{fp}(n,P)$ and $L_{fp}(n,P)$
are compared with data from numerical simulations for the
CGLE parameters $b=0.5,\; c=2.3,\; d=\pm 0.0025$.
The full lines describe stable solutions ($\lambda_{fp}<0$)
while dashed lines correspond to unstable
ones ($\lambda_{fp}>0$).
The theoretical results are seen to be well confirmed by the numerical
simulations especially in the limit of small velocities.
The remaining discrepancies can be explained by the neglect
of the $d$-corrections
in the plane wave quantities ($ k, K,p_{3,4},...$)
and the linearization in $v$.
Note that for fixed period $P$ several
stable solutions can coexist which may (but don't have to)
differ by the phase $2\pi n$ contained in one period (see Fig.6).

The essential influence of the $d$--perturbation
on the solutions is especially apparent for very slowly
solutions (c.f. eq.(\ref{per_sol_2}))
\begin{eqnarray}
  \label{per_sol_8}
  v_{fp} & \approx & -(1/Re(g^* d)) \;
         g_{30}\; e^{-p_{3}'(0)L_{nP}}\; \sin(p_{3}''(0)L_{nP} + g_{40})
\end{eqnarray}
occuring in the limit of small interaction
$e^{-p_{3}'L_{nP}} \ll |d| \ll 1$.
Since one then has $\lambda_{fp}\approx Re(g^* d)$
these solutions are only stable for decelerating
$d$--perturbation.
As a consequence for accelerating $d$
(slowly) uniformly moving solutions exist stably only for
certain periods (see Fig.7) whereas in the opposite case
such solutions occur for arbitrary values of $P$ (see Fig.6).
This fact is well confirmed in our simulations.

Now we turn to the solutions
with oscillating hole velocities.
In our simulations we found such solutions
coexisting (stably) with uniformly moving solutions
but they occurred especially for parameters where eq.(\ref{per_sol_7})
does not have stable fixed points (see Figs.5b,5c).
The solutions are characterized by the occurrence of large
velocities ($v\sim 1$) and interaction strengths
($e^{p_3'(v)L} \sim 1$) during the oscillations.
Presumably they can (in a first approximation)
be identified as stable limit cycles of a two
dimensional dynamical system with
the hole velocity $v$ and the hole--shock
distance $L$ as active variables.
While one has $\dot L = O(\dot v)$
(as in eq.(\ref{per_sol_3}))
$\dot v$, and thus $\dot L$, cannot be neglected now.
Thus from the phase conservation condition one obtains
an evolution equation for $\dot L$ which has to be combined
with the equation for the hole acceleration $\dot v$
yielding the two dimensional dynamical system.
In this description eqs.(\ref{per_sol_6},\ref{per_sol_7})
correspond to the adiabatic elimination of $\dot L$.

The mechanism for the oscillations beyond the core
stability boundary (see Fig.5d) is of different nature.
Here the core instability mode introduces a new degree of freedom.
While for accelerating $d$ this leads to a
destruction of the solutions, in the decelerating case
the interplay of core and acceleration instability leads
to a Hopf bifurcation of the fixed point solutions of
eq.(\ref{per_sol_7}) similar to the situation described
for isolated holes by eq.(\ref{phen}).
This yields the oscillating solutions of Fig.5d.

\section{Concluding Remarks}

The investigation presented in this paper is of general importance
since the CGLE is generally derived from an amplitude expansion
near the threshold of a bifurcation and higher-order perturbations
appear naturally.
Of course there should be other terms of the same
magnitude as the quintic term included in eq.(\ref{eq1}).
However, since the effects described above are closely related
to phase conservation one should expect the other terms to have the same
qualitative consequences and it should be possible to describe a physical
system with a single "effective" higher-order term.
We expect the consequences of our results
to be restricted to (quasi-)1-dimensional systems
in the vicinity of the parameter range where hole
solutions are stable in the (unperturbed) cubic CGLE.
Thus stable, uniformly moving holes should in general not be
observable.
(Note that in the case where the background state carries traveling
waves rather than homogeneous oscillations, the CGLE (\ref{eq1})
has an additional group velocity term.
Motion then has to be defined with respect to a
frame moving with that group velocity.)

The detailed investigation of spatio-temporal chaos presented by
Shraiman et al. \cite{shraiman} for the unperturbed CGLE
is presumably robust with respect to perturbations
because the sensitive parameter range is not
considered there.
A later investigation \cite{chate}
considers the sensitive range,
where the determination of the range of existence of spatio-temporal
chaos becomes questionable in the unperturbed CGLE.
We found that stable hole solutions suppress
spatio-temporal chaos and as a consequence for a stabilizing
perturbation the (upper) boundary of spatio-temporal chaos is simply
given by the stability boundary HS (see Fig.1) of the NB hole solutions,
whereas for destabilizing perturbation spatio-temporal
chaos is also observed further up \cite{PRL}.

Comparing holes in 1d with spirals in 2d one can state
that spirals behave in a rough sense
(apart from topological stability) similar to holes
in the perturbed CGLE with stabilizing (decelerating)
perturbation.
Thus one has a range (which is in 2d very large)
where standing spirals are core stable.
For large values of $|b|$ a core instability occurs,
which is of Hopf type \cite{spiral_inst}.
In contrast to the 1d case the bifurcation is subcritical in 2d
and leads to spatio-temporal chaos.
We also note that the theory of interaction between
localized solutions is similar in 1d and 2d
\cite{inter_2d}.

Transient hole-type solutions were observed experimentally by
Lega et al. \cite{cro} in the (secondary) oscillatory instability in
Rayleigh-B\'enard convection in an annular geometry.
Here one is in a parameter range
where holes are unstable in the cubic CGLE, so that small
perturbations are irrelevant.
Long time stable stationary holes ('1d spirals') were
observed in a quasi-1d chemical
reaction system (CIMA reaction) undergoing a Hopf bifurcation by
Perraud et al. \cite{deKepper}.
The experiments were performed in the vicinity of the
cross-over (codimension-2 point) from the
(spatially homogeneous) Hopf bifurcation
to the spatially periodic, stationary Turing instability.
Simulations
of a reaction-diffusion system (Brusselator) with appropriately
chosen parameters exhibited the hole solutions
(and in addition more complicated
localized solutions with the Turing pattern appearing in the core
region).
It seems likely that for the hole solutions the vicinity of the
codimension-2 point is not important, and that the parameters
correspond
to the stable hole range in the CGLE description with stabilizing
perturbations.

Finally we mention recent experiments by Leweke and Provencal
\cite{Provencal}
where the CGLE is used to describe results of open-flow experiments
on the transitions in the wake of a bluff body in an annular
geometry.
Here the sensitive parameter range is reached and in the observed
amplitude
turbulent states holes should play an important role.

We wish to thank S.Sasa for sending us
data of the core instability line in addition to
those in \cite{sasa}.
One of us (I.A.) wishes to thank the
Alexander-von-Humboldt Stiftung for financial support
and the University of Bayreuth for its hospitality.
Support by the Deutsche Forschungsgemeinschaft
(Kr-690/4, Schwerpunkt
'Strukturbildung in dissipativen kontinuierlichen Systemen:
Experiment und Theorie im quantitativen Vergleich')
is gratefully acknowledged.

\pagebreak
\begin{appendix}
\section{The Nozaki-Bekki Hole Solutions}
\label{NB_a}
Inserting the ansatz (\ref{NB}) into the CGLE one
obtains a polynomial of third order in
$\partial_\zeta\varphi_v(\kappa \zeta)
\equiv K \tanh(\kappa \zeta)$ ($ K:=\kappa /\hat\kappa$).
Requiring that terms at each order cancel separately one
is confronted with 4 complex
equation for 8 real parameters so that a priori
one should only expect a
discrete solution.
The 4 complex equations, however, are not
independent, which leads to the one parameter
family of hole solutions.
In the following we derive the explicit form of the
Nozaki-Bekki
hole solutions in a way which shows this dependence from the
structure of the CGLE and the solutions. (Results are listed
in eqs.(\ref{kappa} -- \ref{elipse})).
Introducing  (\ref{NB}) into the CGLE one obtains
\begin{eqnarray}
\label{CGLE2}
&\left(L_{lin}[v, k,\Omega] - (1+ic)| Z_v |^2\right) Z_v e^{i\varphi_v}=0
\quad, \qquad Z_v=\hat B\partial_\zeta\varphi_v(\kappa\zeta)+\hat A v&
\\
&\mbox{with the linear operator }\quad
L_{lin}[v, k,\Omega]:=
1- k^2+i(\Omega + v^2\hat  k -b  k^2) +  V\partial_\zeta
+(1+ib)\partial_\zeta^2
&
\nonumber
\end{eqnarray}
and  $ V:= v(1 +2i(1+ib)\hat k \equiv v \hat V$).
Using the fact that $L_{lin}$ commutes with the operator
$(-i\hat B\partial_\zeta + v\hat A)$
one easily finds
\begin{eqnarray}
\label{LP}
L_{lin} Z_v e^{i\varphi_v} &=&
L_{lin}  (-i\hat B \partial_\zeta + v\hat A) e^{i\varphi_v}
= \left[\Lambda  Z_v
- \hat B i \partial_\zeta \Lambda\right]  e^{i\varphi_v}
\end{eqnarray}
with the "eigenvalue"
$\Lambda = \Lambda(\partial_\zeta\varphi_v)
:= \exp(- i\varphi_v)L_{lin}   \exp(i\varphi_v)
= z (\partial_\zeta\varphi_v)^2 - iV\partial_\zeta\varphi_v +
1+i(\Omega + v^2\hat  k ) + i(1+ib)(\hat\kappa  K^2 - k^2) $
and  $z:=(1+ib)(i\hat\kappa-1 ) $.
Introducing this into eq.(\ref{CGLE2}) one sees that
$\partial_\zeta \Lambda/Z_v $ has to be a
polynomial of second order in $\partial_\zeta\varphi_v= K \tanh(\kappa\zeta )$
and one finds
\begin{eqnarray}
\label{L}
\Lambda = z\hat B^{-2}  Z_v^2 + const.
\end{eqnarray}
from which follows
\begin{eqnarray}
\label{1}
i\hat V  = 2 z \hat A \hat B^{-1}
\,\, .
\end{eqnarray}
Using eqs.(\ref{LP}) and (\ref{L}) equation (\ref{CGLE2}) reduces to
\begin{eqnarray}
\label{CGLEred}
\Lambda - (1+ic)|Z_v|^2 = 2iz \hat B^{-1}\partial_\zeta Z_v
\end{eqnarray}
which is of second order in $
\partial_\zeta\varphi_v$.
Dividing this equation by $z$
the real part has the form:
\begin{eqnarray}
\hat B^{-2} Re(Z_v^2) - Re(\frac{1+ic}{z}) |Z_v|^2 = const.
\end{eqnarray}
(where the fact that
$\partial_\zeta Z_v=\hat B\partial_{\zeta\zeta} \varphi_v$
is real
 was used).
Noting that
$
Re(Z_v^2)
= (\partial_\zeta\varphi_v+v\hat A')^2 - \hat A''^2
= |Z_v|^2 + const
$
one finds that the equations obtained from the nonconstant
contributions  -- those proportional to
$
(\partial_\zeta\varphi_v)^1$
and $
(\partial_\zeta\varphi_v)^2$ --
 are equivalent.
In this way one can convince oneself that the remaining
equations are indeed not independent.

To get the explicit solutions from eq.(\ref{CGLEred})
one may proceed as follows.
At order $
(\partial_\zeta\varphi_v)^2$
 one obtains a complex equation
\begin{eqnarray}
(1+ic) \hat B^2 = (1 + 2i\hat \kappa) z
\equiv (1 + 2i\hat \kappa) (1+ib)(i\hat\kappa-1)
\end{eqnarray}
which determines $\hat B$ and $\hat \kappa$:
\begin{eqnarray}
\label{kappa}
\hat \kappa = \frac{1}{4(b-c)}\left(3(1+bc)  +
\sqrt{9(1+bc)^2 +8(b-c)^2}\right)
\end{eqnarray}
and
\begin{eqnarray}
\hat B = \sqrt{\frac{3\hat\kappa(1+b^2)}{b-c}}
\,\, .
\end{eqnarray}
{}From order $
\partial_\zeta\varphi_v^1$
contributions of eq.(\ref{CGLEred}) one gets another complex equation
which together with eq.(\ref{1}) yields a set of 3 independent
real equations for
$\hat  k , \hat A' $ and $ \hat A''$.
Solving them one obtains
\begin{eqnarray}
\hat A'
= - \hat B^{-1}
\qquad
\hat A''
=  -2\frac{\hat \kappa}{\hat B}
\qquad
\mbox{and}\quad
\hat  k&=& \left[2(b-c)\right]^{-1}
\end{eqnarray}
where the last formula is identical with the
phase-conservation condition (\ref{topol}).
{}From constant terms
($\sim (\partial_\zeta\varphi_v)^0$)
one can derive the dispersion relation (in the moving frame)
\begin{eqnarray}
\Omega = c-v k +(b-c)( k^2+ K^2)
\end{eqnarray}
and an elliptic relation which relates the two not yet specified
variables $v$ and $\kappa$ (or $ K\equiv\kappa/\hat\kappa$) to each other:
\begin{eqnarray}
\label{elipse}
v^2(\hat  k^2 +  |\hat A|^2) + K^2(1 +\hat B^2)=1
\,\, .
\end{eqnarray}
Thus one finds indeed a one parameter family of moving hole solutions which can
be labeled e.g.  by their velocity.
{}From eq.(\ref{elipse}) one finds that $\kappa$ becomes zero at
$v=v_{max}\equiv
\left[\hat  k^2 +  |\hat A|^2\right]^{-1}$.

\section{Asymptotic Analysis of Nozaki--Bekki Hole Solutions}
\label{app_pws}
%
%
In the limit $ \zeta \rightarrow \pm\infty $ the NB hole solutions
(\ref{NB}) become plane waves
with wavenumber $ q_{1/2} :=  k \pm  K $
(see eq.(\ref{q}))
and frequency $ \omega(q_{1/2}) = c + (b-c)q_{1/2}^2 $
(see eq.(\ref{eq2})).
The perturbational equation for these plane waves in a
coordinate system moving with the velocity $v$ of the
hole core reads
\begin{eqnarray}
  \label{a204}
  \partial_t W   & = &
  (1 \! + \! ib) \; \partial_{\zeta \zeta} W   \; + \;
  (v + 2 i q_{1/2} (1 \! + \! ib)) \; \partial_{\zeta} W  \; - \;
  (1 \! + \! ic) (1 \! - q_{1/2}^2) \; ( W + W^* )
\end{eqnarray}
This equation coincides with the asymptotic
($|\kappa\zeta|\gg 1$) perturbational equation
(\ref{L_eck}) with $L_v$ from eq.(\ref{Lv})
(after transforming away an (irrelevant) fixed phase factor
 $W\to W e^{-i arg(v \hat A \pm  K \hat B)}$).
Equation (\ref{a204}) can be solved by an exponential ansatz
\begin{eqnarray}
  \label{a205}
   W = z e^{p \zeta + \lambda t} + z_* e^{p^* \zeta + \lambda^* t}
  \qquad \qquad
  \mbox{with} \quad p,\lambda,z,z_* \; \mbox{complex}
\end{eqnarray}
Inserting this into eq.(\ref{a204}) yields two complex linear equations
for the constants $ z,z_*^* $ whose solvability condition leads to
characteristic equation
\begin{eqnarray}
  \label{a206}
  p^4 (1+b^2) \; + \;
  p^3 \; 2v  \; & + & \;
  p^2 \left( 4 q_{1/2}^2 + (v-2bq_{1/2})^2 -
                  2(1+bc)(1-q_{1/2}^2) - 2\lambda \right)
    \nonumber \\
  & + & \;p \; \left( (4(b-c)q_{1/2} - 2v)(1-q_{1/2}^2)
                                + (4bq_{1/2}-2v)\lambda \right) \; + \;
  \left( \lambda^2 + 2\lambda (1-q_{1/2}^2) \right)
  \; = \; 0
\end{eqnarray}
The four roots $ p_i = p_i(v,\lambda) $ describe the possible
asymptotic spatial behavior of perturbations $W$
with growth rate $ \lambda $ in the coordinate system co-moving with
the velocity $ v $ of the hole core.
In particular the exponents $ p_i = p_i(v) $ describing
stationary perturbations are obtained for
$ \lambda=0 $.
For $\zeta \to + \infty$ one finds
\begin{eqnarray}
  \label{a207}
  p_1 & = & 0 , \qquad  p_2 = - 2 \kappa  \nonumber \\
  p_{3/4} & = &
     \kappa - \frac{v}{1+b^2} \; \pm \;
     \sqrt{ - 4q_{1}^2 - 3 \kappa^2
            + 2(1-q_{1}^2) \frac{1+bc}{1+b^2}
            + \frac{4bq_{1}v + 2\kappa v - v^2}{1+b^2}
            + \frac{v^2}{(1+b^2)^2} }
\end{eqnarray}
For $ v=0 $ one has $ q_1 = -q_2 $ and the expression
for $ p_{3/4} $ simplifies
\begin{eqnarray}
  \label{a208}
  p_{3/4} & = & \kappa \pm \sqrt{ \kappa^2 - 6 q_1^2 }
\end{eqnarray}
The exponents $p_i$ for $\zeta\to -\infty$
can be found from eqs.(\ref{a207},\ref{a208})
by the replacement $q_1 \to q_2$ and reversal of the overall signs.
In the following we will for simplicity only treat the case
$\zeta\to+\infty$.
The exponents $ p_1 $ and $ p_2 $ describe
the asymptotics of the two
neutral modes eqs.(\ref{phi_rot},\ref{phi_tr}).
The exponents $ p_{3/4} $ being real (monotonic case)
or complex conjugated (oscillatory case)
always satisfy
\begin{eqnarray}
  \label{a211}
  Re(p_{3/4}) & > & 0 \; ,
\end{eqnarray}
which shows that two fundamental modes of $ L^{NB}_v $
are exponentially growing in space.
For each $ p_i $ the constants $ z(p_i),z_*(p_i) $ can be found
from the eigenvector equation. One finds
\begin{eqnarray}
  \label{a212}
  z_* & = & \eta z^*
  \nonumber \\
  \mbox{where} \quad \eta(p_i) & := &
    \left( \frac{p_i^2 (1+ib) + p_i (v-2bq_1+2iq_1)} {(1+ic)(1-q_1^2)} -1
\right)^*
\end{eqnarray}
For real $p_i$ (which covers the exponents $ p_1, p_2 $
and in the monotonic case also $ p_3, p_4 $) one has
$ |\eta(p_i)| = 1 $
and the asymptotic behavior of the stationary perturbation
associated with one of the exponents $ p_i $ is given by
\begin{eqnarray}
  \label{a214}
  W & \sim &  e^{p_{i} \zeta + i\phi_i}
  \quad \mbox{where} \;\; \phi_i := \frac{1}{2} arg(\eta(p_i))
  \qquad \quad \mbox{for} \; \zeta \rightarrow \infty
  \qquad \quad (p_i \; \mbox{real})
\end{eqnarray}
In the case $ p_3 = p_4^* $ complex $ W $ behaves asymptotically like
\begin{eqnarray}
  \label{a215}
  W & = &   z e^{p_{3} \zeta }
            + \eta(p_{3}) (z e^{p_{3} \zeta })^*
  \qquad \mbox{for} \; \zeta \rightarrow \infty \qquad
  (p_{3} \; \mbox{complex})
\end{eqnarray}
The complex quantity $z$ gives
the two growing fundamental modes in the perturbation $ W $.
We note that depending on the velocity $v$ of the hole the
behavior of the growing perturbations $\sim e^{p_{3/4}\zeta}$
may be quite different on both wings;
in particular it may be oscillatory on one and
monotonic on the other wing.

Now we turn to the situation of a small nonzero
growth rate ($ |\lambda|\ll 1 $).
Then the exponents from eq.(\ref{a207}) are slightly changed.
In particular the exponent $ p_1=p_1(v,\lambda) $
describing the outer asymptotics of localized modes
bifurcating through the translation mode $\Phi_{tr}$
(e.g. the core instability mode)
becomes near their bifurcation
\begin{eqnarray}
  \label{a216}
  p_1 \approx \frac{\lambda}{v-2|q_1(b-c)|}
  \qquad \qquad \mbox{for} \quad |\lambda| \ll 1
\end{eqnarray}

Performing a similar analysis of asymptotic plane wave states in the
perturbed CGLE ($0\neq |d |\ll 1$) most of the foregoing expressions
receive $d$--corrections.
However, since translational (and rotational) invariance are preserved
by the perturbed equation, the stationary ($\lambda =0$)
exponent $p_1=0$ remains unchanged
and also eq.(\ref{a216}) for small $\lambda$
remains valid at lowest order in $d$.
Therefore eq.(\ref{a216}) describes
(besides the core instability mode in the cubic CGLE)
also the acceleration mode in the perturbed CGLE.

\section{Core Instability}
\label{app_core}
In this section we present the solution of the boundary value problem
eqs.(\ref{solv_core_2},\ref{bound_f})
defining the core instability line utilizing the ansatz (\ref{core_ansatz}).
{}From the relations
\begin{eqnarray}
  \label{a31}
  \hat L \; (\zeta \cosh^{-2n}(\kappa \zeta))
    & = &
    \cosh^{-2n}(\kappa \zeta) \;
    \left( N_n \zeta - M_n^c \zeta \cosh^{-2}(\kappa \zeta)
     - R_n \tanh(\kappa \zeta) \right)
      \\
  \hat L \; (\tanh(\kappa \zeta) \cosh^{-2n}(\kappa \zeta))
    & = &
    \cosh^{-2n}(\kappa \zeta) \;
    \left( N_n \tanh(\kappa \zeta)
     - M_n^d \tanh(\kappa \zeta) \cosh^{-2}(\kappa \zeta) \right)
      \\
  \mbox{with} & &
    N_n := 4 n^2 \kappa^2 + 12 n \kappa^2 + 8 \kappa^2 + 6  K^2 ,
    \quad \;
    R_n := 4 \kappa n + 6 \kappa
      \\
  & &
    M_n^c := 4 n^2 \kappa^2 + 14 n \kappa^2 + 12 \kappa^2 + 6  K^2 ,
    \quad
    M_n^d := 4 n^2 \kappa^2 + 18 n \kappa^2 + 20 \kappa^2 + 6  K^2 ,
\end{eqnarray}
follows, that when inserting (\ref{core_ansatz})
into eq.(\ref{solv_core_2}),
one can in a first step recursively determine the coefficients $ c_m $
and then in the next step the coefficients
$ d_m $ from the $ c_m $.
In summary one has
\begin{eqnarray}
   & &
     c_{-1} = 1 , \qquad
     c_0    = \frac{\Delta_1}{N_0} ,
     \\
   & &
  \label{a32}
     c_m    = \frac{\Delta_1}{N_m} \prod\limits_{n=0}^{m-1} \frac{M_n^c}{N_n}
     \\
   & &
     d_{-1} = \frac{\Delta_2}{N_{-1}} ,
     \\
   & &
  \label{a33}
     d_m    = \frac{1}{N_m} \{ d_{m-1} M_{m-1}^d + c_m R_m \}
            = \frac{1}{N_m} \;
              \left( \prod\limits_{n=0}^{m} \frac{M_{n-1}^d}{N_{n-1}} \right)
              \left\{ \Delta_2 + \Delta_1 \frac{N_{-1}}{M_{-1}^d}
                    \left( \frac{R_0}{N_0} +
                    \sum\limits_{n=1}^m \frac{R_n}{N_n}
                        \prod\limits_{k=0}^{n-1} \frac{M_k^c}{M_k^d}
                    \right)
              \right\}
\end{eqnarray}
where
\begin{eqnarray}
   &
   \Delta_1 := 2 \kappa^2 + 3 b \kappa  K ,
   \qquad
   \Delta_2 := \frac{ K}{1- K^2}
   ( 3 b + 3 b  K^2 - 14 \hat \kappa  K^2 + 2 \hat \kappa )
   &
\end{eqnarray}
{}From eq.(\ref{a32}) follows that for large $ m $ the coefficients $ c_m $
behave like
\begin{eqnarray}
     c_m \; = \; \frac{\Delta_1}{N_m} \prod\limits_{n=0}^{m-1}
\frac{M_n^c}{N_n}
         \; \sim \; \frac{1}{m^2} \prod\limits_{n=0}^{m-1}
                        \frac{1+\frac{7}{2n}}{1+\frac{3}{n}}
         \; \sim \; m^{-3/2} \quad \mbox{for} \; m \rightarrow \infty
\end{eqnarray}
Therefore the first sum in
(\ref{core_ansatz}) always converges uniformly towards a
smooth function.
Because of
\begin{eqnarray}
     \frac{1}{N_m} \prod\limits_{n=0}^{m-1} \frac{M_n^d}{N_n}
     \; \sim \; \frac{1}{m^2} \prod\limits_{n=0}^{m-1}
                        \frac{1+\frac{9}{2n}}{1+\frac{3}{n}}
     \; \sim \; m^{-1/2} \quad \mbox{for} \; m \rightarrow \infty
\end{eqnarray}
the coefficients $ d_m $ in eq.(\ref{a33}) may behave like
$ \; \sim m^{-3/2} $ or $ \; \sim m^{-1/2} $ depending on
whether the curly bracket in eq.(\ref{a33}),
which behaves like
$ \; const + O(1/m) \; $,
gives zero in the limit $ \; m\rightarrow \infty $ or not.
In the latter case the second sum in the ansatz eq.(\ref{core_ansatz})
converges (pointwise) towards a function with a finite step
at $ \zeta = 0 $ whereas in the first case the limiting function
is smooth everywhere.
Therefore one has to adjust $ (b,c) $ with $ b=b(c) $ in a way that
the curly bracket in eq.(\ref{a33})
vanishes for $ \; m\rightarrow \infty $.
This determines the core instability line.
Resubstitution of $ \;  N_m , M_m^c , ... $ into
the curly bracket of eq.(\ref{a33}) yields
the final expression (\ref{core_line}).

\section{Acceleration Instability in the Nonlinear Schroedinger Limit}
\label{nls_acc_inst}
In this section we present a method to treat the acceleration
instability in the NLS-limit ($b,c\to\infty$)
by making use of the formalism introduced in
subsections \ref{matching},\ref{stat_bif},\ref{acc_inst}.
We calculate the acceleration by projecting the (symmetric)
inhomogeneity of eq.(\ref{acc_inst_6}) onto
the symmetric neutral mode $ \Phi_s $ of the adjoint operator
\footnote{In this section we drop the
          indices '0' and 'NB' to simplify the notation.}
$ L^{\dagger} := L^{NB \; \dagger}_0 $.
The scalar product vanishes only for suitably adjusted
parameter $ \frac{\dot v}{v} $ in the inhomogeneity
and this ensures the solvability
of eqs.(\ref{acc_inst_6},\ref{bound_inner}).

Using the scalar product
\begin{eqnarray}
  \label{acc_nls_01}
  \langle U , W \rangle := \int_{-\infty}^{\infty} Re(U^* W)dx
\end{eqnarray}
the adjoint problem is
\begin{eqnarray}
  \label{acc_nls_02}
  L^{\dagger} \Phi_s  = 0
  \qquad \mbox{with} \qquad
  |\Phi_s| \to 0
  \qquad \mbox{for} \;\;
  \zeta \to \pm\infty
\end{eqnarray}
\begin{eqnarray*}
  L^{\dagger} \Phi_s \; = \;
  (1 \! - \! ib) \; \partial_{\zeta \zeta} \Phi_s   \; + \;
  (2 i  K (1 \! - \! ib) \tanh (\kappa \zeta)) \;
            \partial_{\zeta} \Phi_s    \; +  \;
  (1 - i \Omega + (1 \! - \! ib) i  K \kappa ) \frac{1}{\cosh^2(\kappa \zeta)}
             \;  \Phi_s    \; - \;
  \qquad \qquad \qquad \nonumber    \\
  \qquad \qquad \qquad
  - \; (1 \! - \! ic) (1 \! - \!  K^2) \tanh^2 (\kappa \zeta) \; \Phi_s
  - \; (1 \! + \! ic) (1 \! - \!  K^2) \tanh^2 (\kappa \zeta) \; \Phi_s^*
\end{eqnarray*}
In analogy to Appendix \ref{app_pws} an investigation of the asymptotics
of the fundamental modes of $ L^{\dagger} $
shows that for $ | \kappa \zeta | \gg 1  $
they behave like $ \; \sim e^{p_i |\zeta|} $.
For $\zeta\to\infty$ the exponents $ p_i $ are
\begin{eqnarray}
  \label{acc_nls_03}
  p_1 = 0, \qquad p_2 = +2\kappa,
  \qquad p_{3/4} = -\kappa \pm \sqrt{\kappa^2-6 K^2}
\end{eqnarray}
These are just the exponents of $ L $ with reversed sign
(c.f. eqs.(\ref{a207},\ref{a208})).
The two decaying exponents $ p_{3/4} $ describe
the desired symmetric mode $ \Phi_s $
in the outer region
($ | \kappa \zeta |
  \gg 1 $)
of the asymptotic matching procedure.
In the NLS limit one finds
with the help of the relations of App.\ref{NB_a}
\begin{eqnarray}
  \label{acc_nls_04}
  p_{3/4} = -\kappa \pm \kappa \left( 1 - \frac{4(b-c)^2}{3b^2c^2}
  + O(b^{-4}) \right)
\end{eqnarray}
Both exponents are real ('monotonic range')
and $ p_3 \to 0 $ for $b,c\to\infty$.
The contribution to $\Phi_s$ proportional to
$ e^{p_3 \zeta} $ has a prefactor $z_3$ with
\begin{eqnarray}
  \label{acc_nls_05}
  \frac{Im(z_3)}{Re(z_3)} =
   - \left( \frac{1}{c} +  O(b^{-3}) \right)
\end{eqnarray}
as can be calculated in analogy to (\ref{a214}).
The inner region is defined by
$ | p_3 \zeta | \cong | \kappa \zeta \frac{4(b-c)^2}{3b^2c^2}| \ll 1 \; $
which ensures a finite overlap
in the NLS limit.
Here we expand the differential equation (\ref{acc_nls_02}) in
terms of $ 1/b $
\begin{eqnarray}
  \label{acc_nls_06}
     L & = & b \left( L^{(0)} + b^{-1} L^{(1)} + O(b^{-2}) \right)
     \nonumber \\
     L^{\dagger} & = & b \left( L^{(0)\dagger}
                          + b^{-1} L^{(1)\dagger} + O(b^{-2}) \right)
     \nonumber \\
     \Phi_s & = & \Phi_s^{(0)} + b^{-1} \Phi_s^{(1)} + O(b^{-2})
\end{eqnarray}
At order $ b^1 $ eq.(\ref{acc_nls_02}) reads
\begin{eqnarray}
  \label{acc_nls_07}
     \underline{b^1}: & \quad &  L^{(0)\dagger}  \Phi_s^{(0)}
     \; = \;
     -i \partial_{\zeta \zeta} \Phi_s^{(0)} \; + \;
     \frac{ic}{b} (2 \tanh^2(\kappa \zeta) -1) \Phi_s^{(0)} \; - \;
     \frac{ic}{b} \tanh^2(\kappa \zeta) \Phi_s^{(0) *} \; = \; 0
\end{eqnarray}
from which one gets $ \Phi_s^{(0)} = i/\cosh^2(\kappa \zeta) $.
At order $ b^0 $ eq.(\ref{acc_nls_02}) yields
\begin{eqnarray}
  \label{acc_nls_08}
     \underline{b^0}:  \quad
      L^{(0)\dagger} \Phi_s^{(1)} & = &
      -  L^{(1)\dagger} \Phi_s^{(0)}
       \nonumber \\
   & = & i \; \left( \tanh^2(\kappa \zeta) (8-28\kappa^2)
             + (-4+8\kappa^2) \right)/(3\cosh^2(\kappa \zeta))
\end{eqnarray}
Using the full fundamental system of $L^{(0)\dagger}$
(irrespective of boundary conditions)
\begin{eqnarray}
  \label{acc_nls_09}
  i/\cosh^2(\kappa \zeta) & \qquad  \qquad &
     \frac{i}{4\kappa} \cosh(\kappa \zeta) \sinh(\kappa \zeta)
     + \frac{3i}{8\kappa} \tanh(\kappa \zeta)
     + \frac{3i}{8} \frac{\zeta}{\cosh^2(\kappa \zeta)}
     \nonumber \\
  \tanh(\kappa \zeta)
     & \qquad  \qquad &
     1 - \kappa \zeta \tanh(\kappa \zeta)
\end{eqnarray}
one can construct the bounded solution
\begin{eqnarray}
  \label{acc_nls_10}
   \Phi_s^{(1)} & = &
      \frac{2-7\kappa^2}{3\kappa^2} \tanh^2(\kappa \zeta) + 1
\end{eqnarray}
Combining $ \Phi_s^{(0)} $ and $ \Phi_s^{(1)} $
and taking the limit $\zeta\to\infty$ in the terms $ O(b^{-1}) $
one finds the outer behavior of the inner solution
\begin{eqnarray}
  \label{acc_nls_11}
   \Phi_s & = &
      i/\cosh^2(\kappa \zeta) \; + \;
      \frac{2-4\kappa^2}{3b\kappa^2} \; + \; O(b^{-2})
\end{eqnarray}
This expression has to be matched to the outer exponential solution.
By using eqs.(\ref{acc_nls_03}-\ref{acc_nls_05},\ref{acc_nls_11})
one finally gets
\begin{eqnarray}
  \label{acc_nls_12}
   \Phi_s & = &
   \frac{i}{\cosh^2(\kappa \zeta)} \left(1+O(b^{-1})\right)
   \; + \;
   \frac{2-4\kappa^2}{3b\kappa^2} \; e^{p_3 |\zeta|}
    \left(1 - i \left(\frac{1}{c} + O(b^{-3}) \right)
         \right) (1+O(b^{-1}))
\end{eqnarray}

Now we turn to the inhomogeneity of eq.(\ref{acc_inst_6}).
For arbitrary $b,c$ the terms $\Delta\Omega, \; W^{(1)}$
occuring here can be found from the following ansatz
describing the standing hole of the perturbed cubic CGLE
in the inner region
\begin{eqnarray}
  \label{acc_nls_13}
  \left(\mu_2\tanh(\mu_1x) + \mu_3\tanh^3(\mu_1x) \right) \;
  \exp\left(i\mu_4\ln\cosh(\mu_1x) + i\mu_5\tanh^2(\mu_1x) - i\mu_6 t \right)
  \qquad \mbox{with} \; \mu_i \; \mbox{real}
\end{eqnarray}
Taking the NLS limit of the inhomogeneity
in eq.(\ref{acc_inst_6}) and projecting it
onto the neutral mode $\Phi_s$ then yields the acceleration
\begin{eqnarray}
  \label{acc_nls_14}
  \frac{\dot v}{v}
     \left( \frac{3b^4c}{2(b-c)^3} \left(1+O(b^{-1})\right) + \frac{8}{3}
            - \frac{b}{c} - \frac{2b}{b-c} + O(b^{-1}) \right)
  & = & \frac{16}{15} Re(d)
\end{eqnarray}
As stated in Sec.\ref{stat_bif} the acceleration
$\dot v/v$ diverges at the boundary of the core
instability (eq.(\ref{core_lim})).

\end{appendix}

\pagebreak

\pagebreak

	{\bf Figure Captions}
	\newcounter{fig}

\begin{list}{\bf Figure \arabic{fig}:}{\usecounter{fig}}

\item Stability diagram for the 1D CGLE.
Waves emitted by the standing hole
become convectively unstable below the
dashed curve EH (Eckhaus instability)
and absolutely unstable below the solid curve HS.
Outside the region bounded by the solid curve CS
the core of the standing hole becomes unstable
(Hopf bifurcation for (real) $d<0$).
The CS line is given by eq.(\ref{core_line}).
Stable standing holes can be found above HS and
below CS (upper branch) for $d\leq 0$.
The dashed dotted line MO gives the boundary between monotonic
(above) and oscillatory (below) interaction
for standing holes (see eq.(\ref{int_1})).

\item Acceleration instability in the perturbed (cubic) CGLE.
Comparison of the reduced acceleration $\partial_t v /v$
from theory (full line) and simulations (squares)
for $b=0.5, \; c=2.0, \; |d|=0.002$
and varying phase $arg(d)$.
The theoretical curve was found from adjusting
(numerically) the parameters $\dot v$ and $\Delta \Omega$ in
eq.(\ref{acc_inst_3}) leading to
$\dot v/v \; = \; Re(0.6572 \; e^{0.6690 i}\ d)$,
see eq.(\ref{acc_inst_7}).

\item Acceleration $\dot v$ of a standing hole in
the presence of a shock at distance $L$
in the cubic CGLE ($d=0$).
Comparison of simulations (squares) and theory.
The parameters $r_3,r_4,z$ in the boundary condition
eqs.(\ref{int_1},\ref{int_2}) were obtained from the
(numerical) nonlinear shock solution (full lines)
or from the analytic approximation
eq.(\ref{int_3}) (dashed lines).
(a) monotonic range: $b=1.0, c=3.5$;
full line:   $\dot v = 3.395 \;e^{-p_4 L}$,
dashed line: $\dot v = 4.255 \;e^{-p_4 L}$
with $p_4=\kappa-\sqrt{\kappa^2-6q^2} \approx 0.4877$
(see eq.(\ref{a208})).
(b) oscillatory range: $b=0.5, c=2.3$;
full line:   $\dot v = 4.361 \;e^{-Re(p_3) L} \sin(Im(p_3) L -1.460)$;
dashed line: $\dot v = 8.823 \;e^{-Re(p_3) L} \sin(Im(p_3) L -0.874)$;
with $p_3=\kappa+\sqrt{\kappa^2-6q^2} \approx 0.8896+0.5940i$
(see eq.(\ref{a208})).

\item Snapshot of the modulus $|A|=|A(x)|$ of a stable
uniformly moving hole--shock pair in a simulation for
$b=0.5,\; c=2.3,\; d=+0.0025$.
The solution is space periodic with period $P=48.4$.

\item Simulations showing the velocity of the hole
center $v=v(t)$ of interacting
hole--shock pairs (periodic boundary conditions).
In Figs.(a),(b),(c) the CGLE parameters were
$b=0.5, c=2.3, d=+0.0025$
(i.e. far away from the core instability line).
(a) relaxation into a constantly moving solution
    for period $P=48.4$.
(b) selected final state with (almost) harmonic oscillating
    velocity for period $P=37.0$.
(c) selected final state with anharmonic oscillating
    velocity for period $P=40.0$.
(d) CGLE parameters $b=0.21, c=1.3,d=-0.005$
    (near the core instability line);
    state with oscillating velocity
    (including change of the direction)
    for period $P=50$.

\item Interacting hole--shock pairs.
Velocity $v=v(n,P)$ and bound state distance $L=L(n,P)$
are plotted as function of the period $P$
for different phases $2n\pi$ contained in one period
for the CGLE parameters
$b=0.5, c=2.3, d=-0.0025$.
Full lines show stable ($\lambda_{fp}<0$)
and dashed lines unstable ($\lambda_{fp}>0$)
uniformly moving solutions which are predicted
(fixed points) by eq.(\ref{per_sol_7}).
They are compared with (stable) uniformly moving solutions
observed in simulations (X).
For larger velocities there are more stationary and oscillatory
solutions, which have not been included since the analysis is not
applicable in that range.

\item
Like Fig.6, but
with CGLE parameters
$b=0.5, c=2.3, d=+0.0025$.

\end{list}

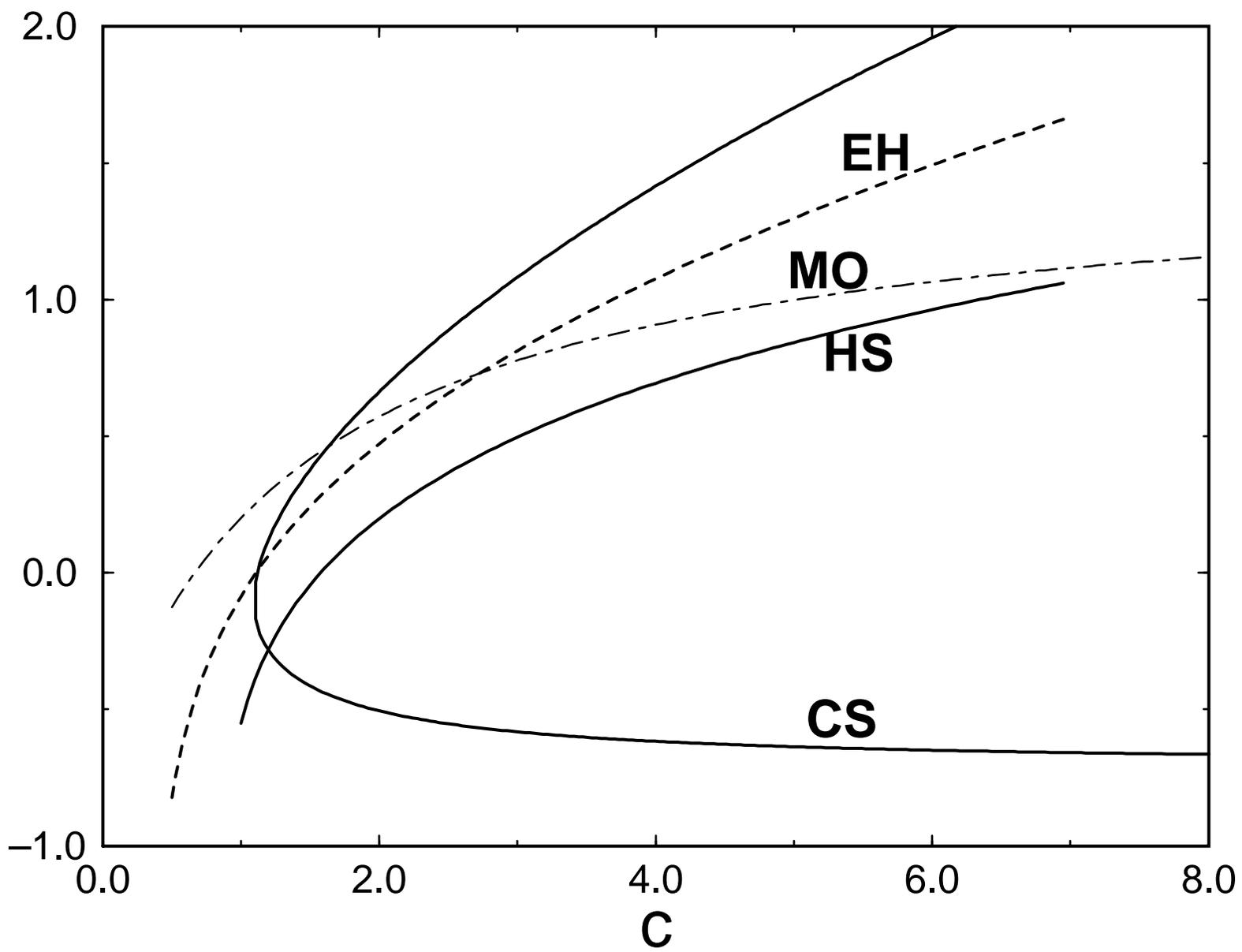
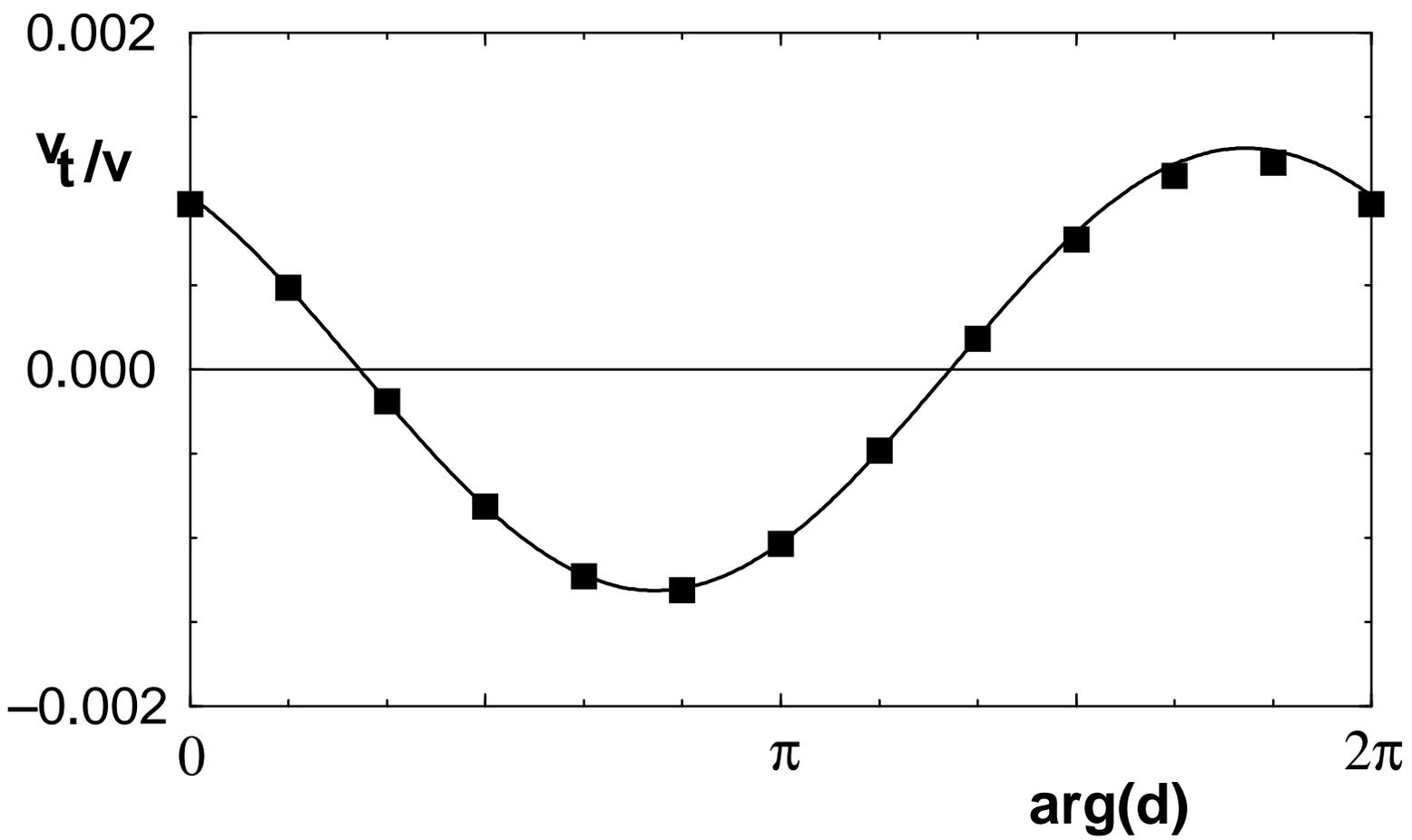
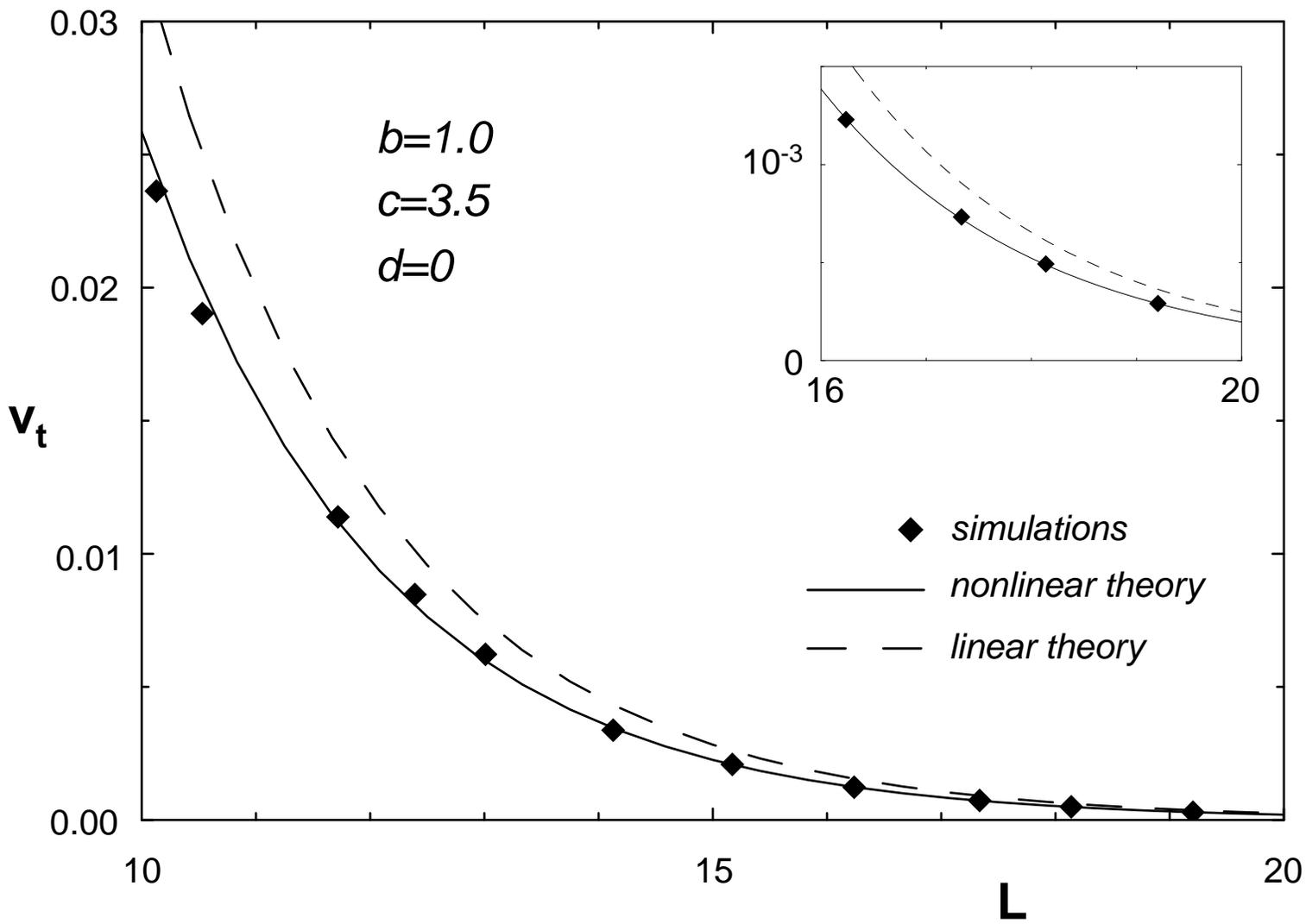
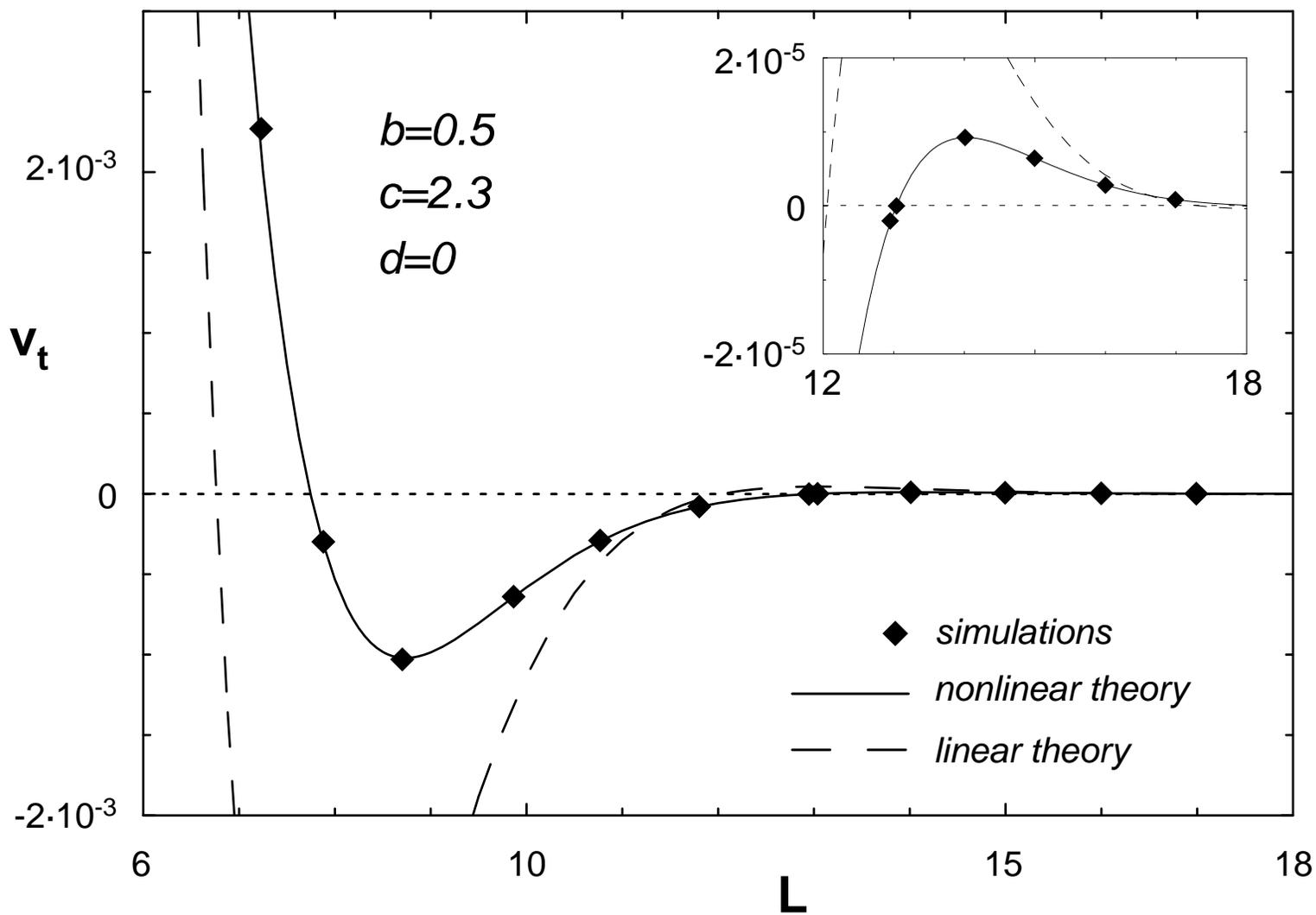
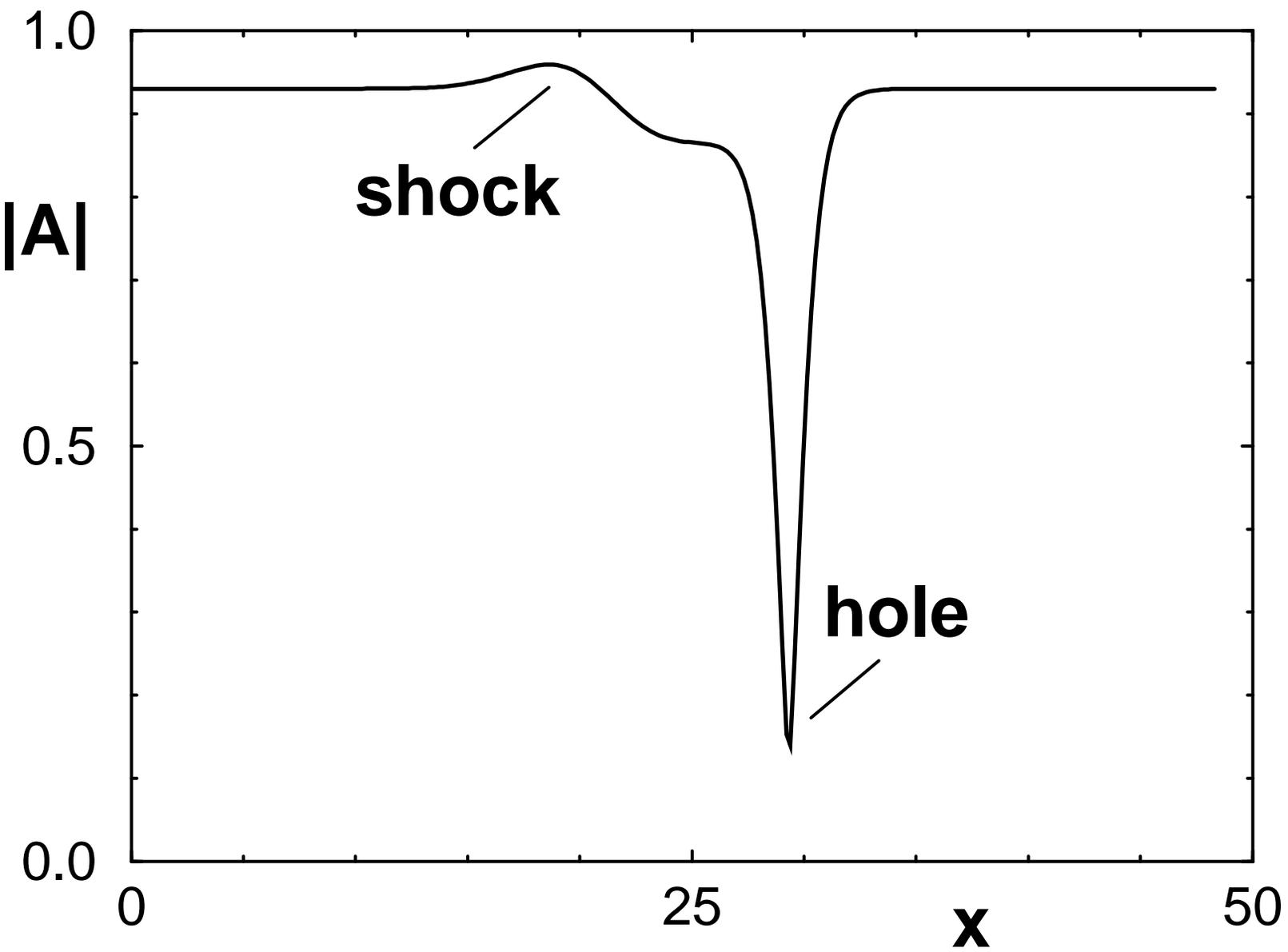
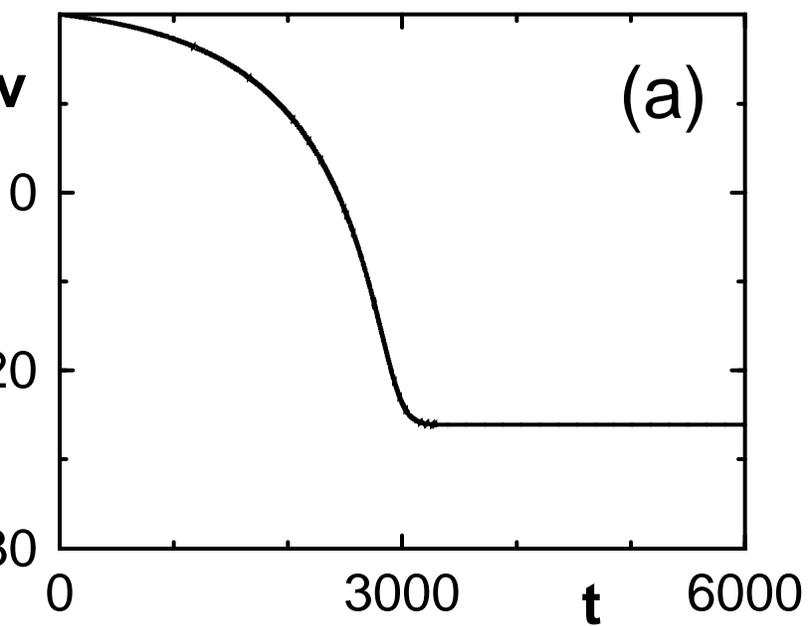
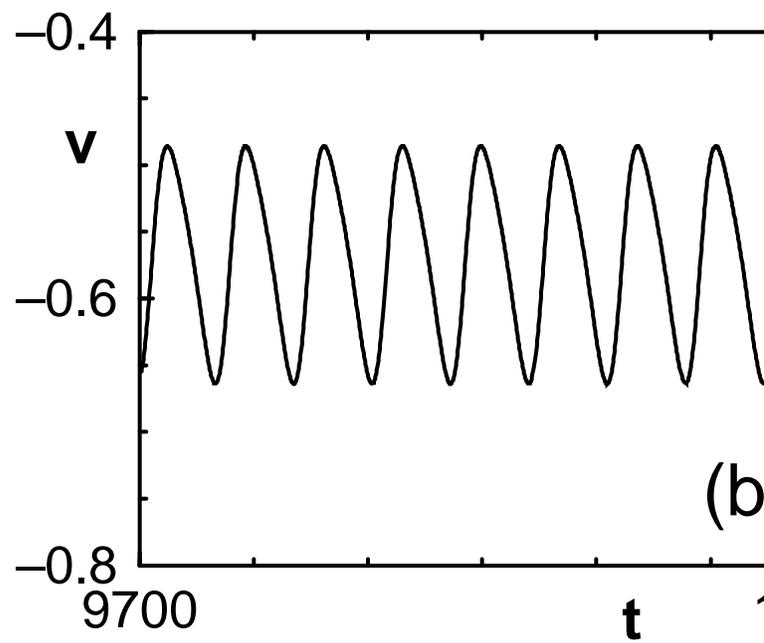
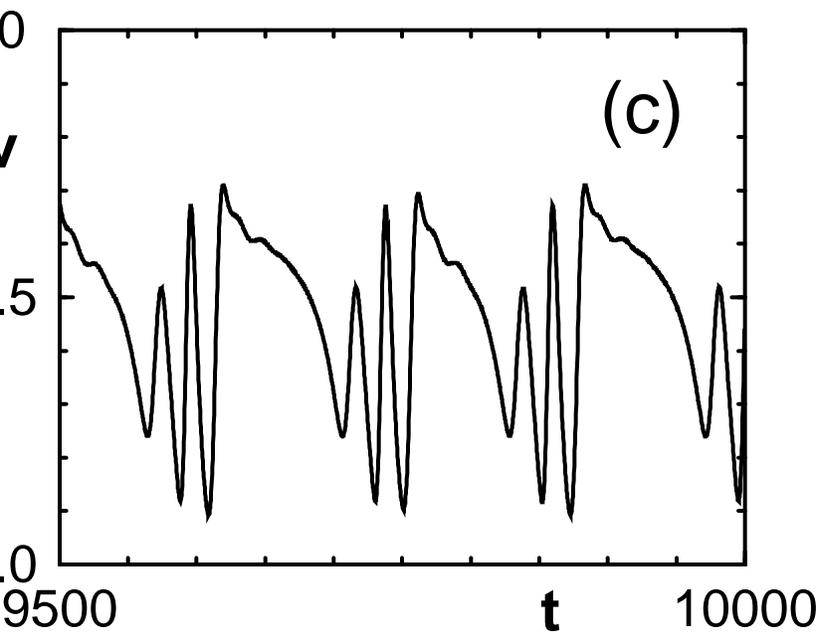
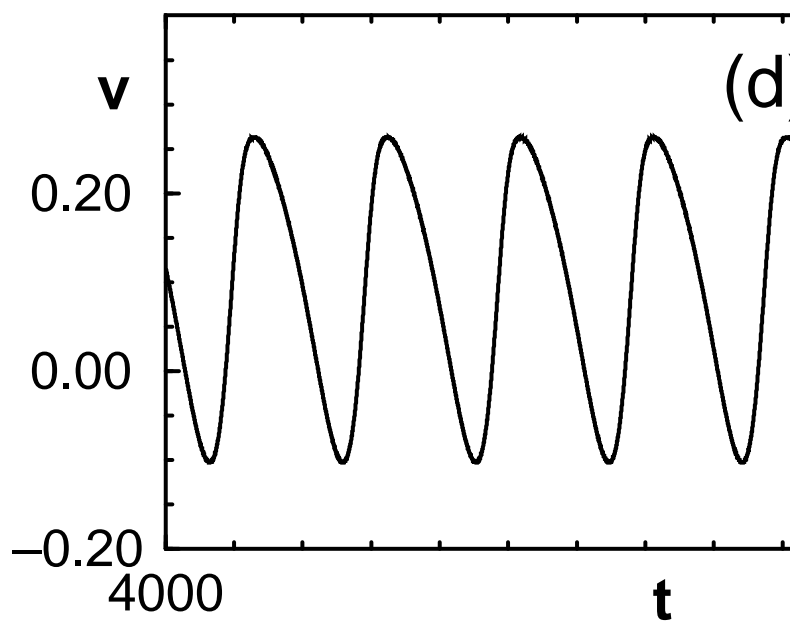
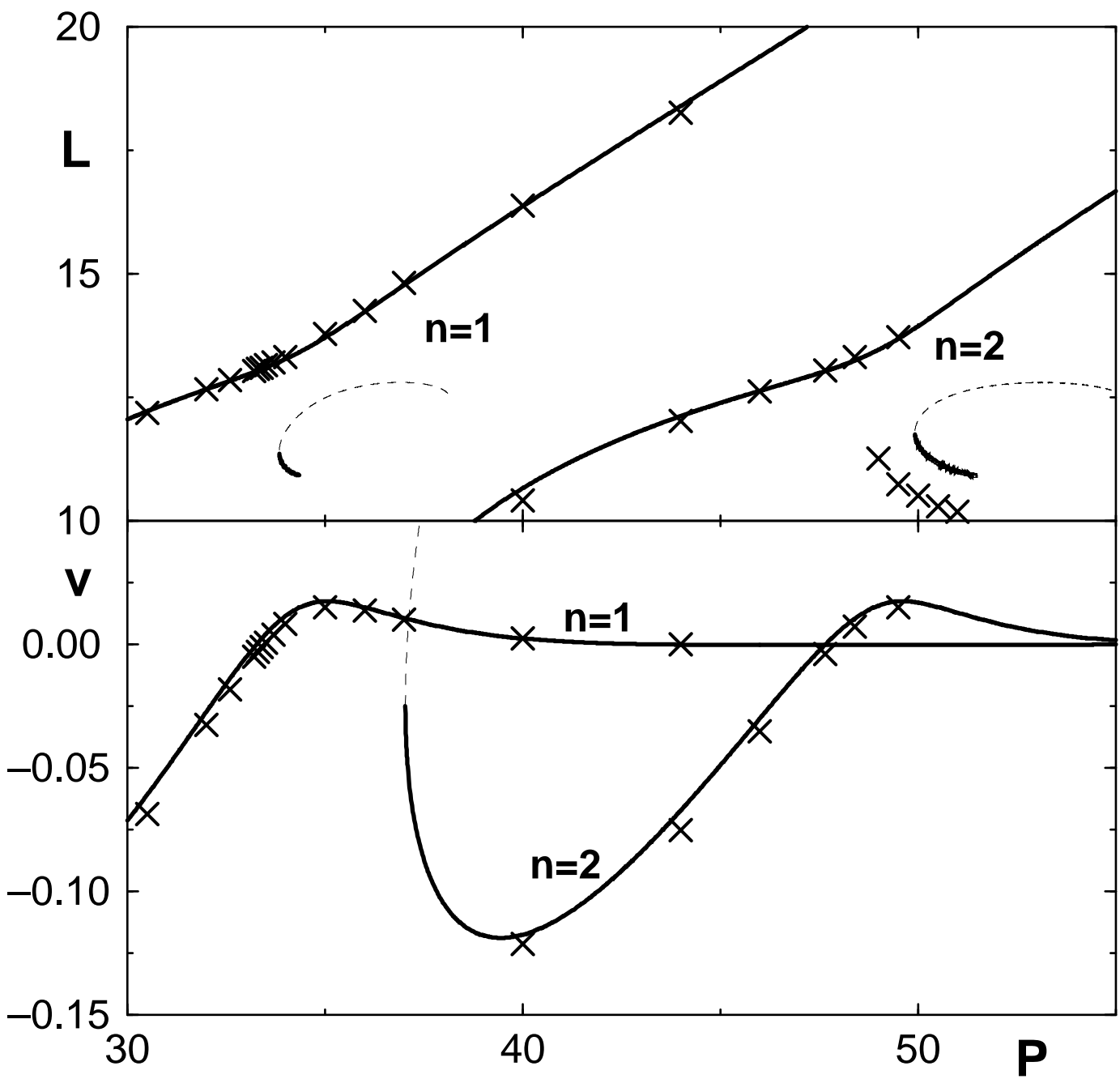
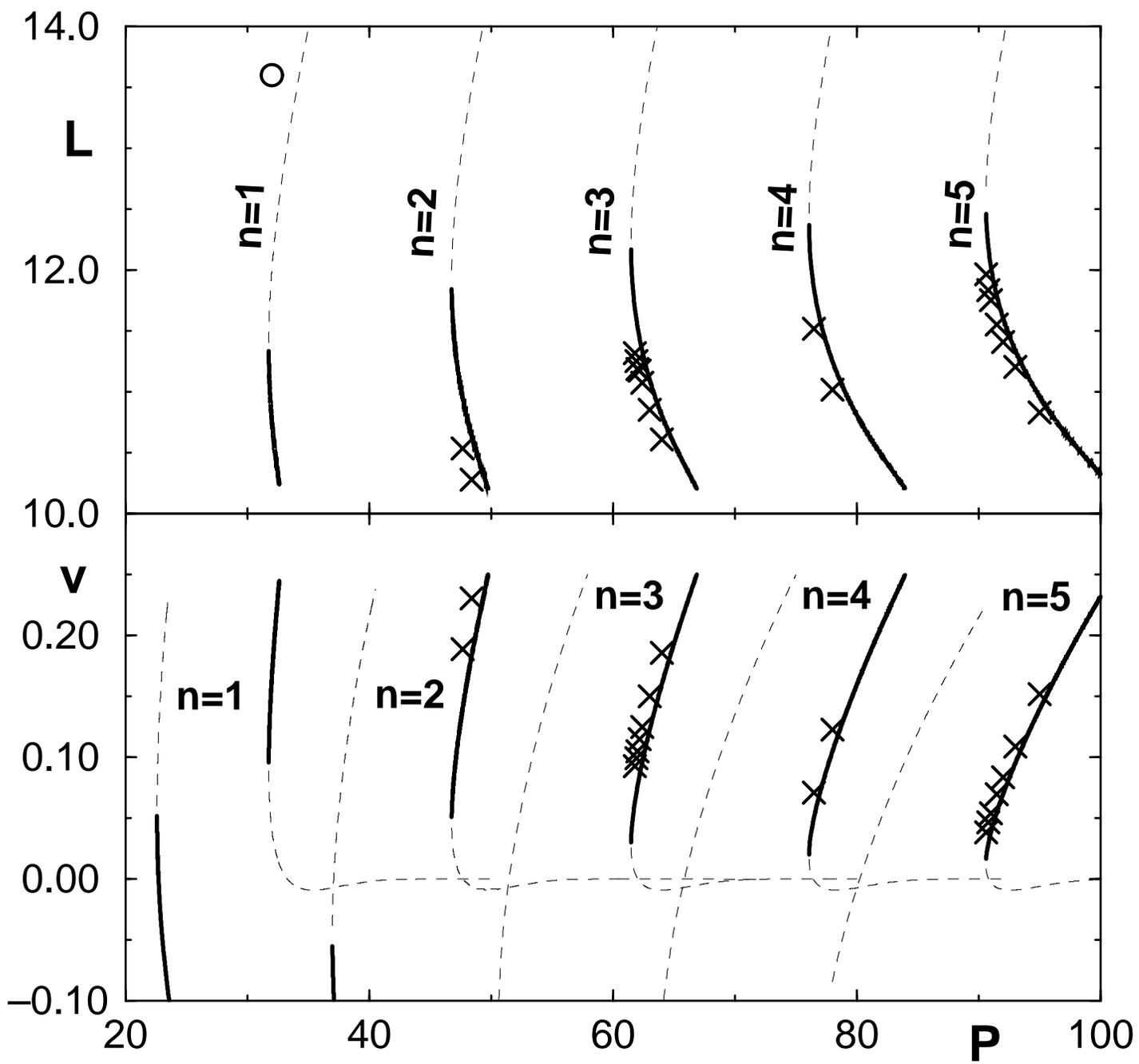
\end{document}